\documentclass[prb,aps,twocolumn,superscriptaddress,floatfix]{revtex4-2}
\usepackage[pdfencoding=auto, psdextra, colorlinks,bookmarks=true,citecolor=blue,linkcolor=red,urlcolor=blue]{hyperref}

\usepackage{graphicx,amsfonts,amssymb,amsmath,dsfont}
\usepackage{multirow}
\usepackage{times}
\usepackage{soul}

\usepackage{young}
\usepackage{youngtab}

\graphicspath{{./}}

\newcommand{\beq}{\begin{equation}}
\newcommand{\eeq}{\end{equation}}
\newcommand{\beqa}{\begin{eqnarray}}
\newcommand{\eeqa}{\end{eqnarray}}
\newcommand{\ket} [1] {\vert #1 \rangle}
\newcommand{\bra} [1] {\langle #1 \vert}

\def\bra#1{\langle#1\vert}
\def\ket#1{\vert#1\rangle}

\def\Longarrow{\protect\@lra}
\def\@lra{\relbar\joinrel\relbar\joinrel\relbar\joinrel%
          \relbar\joinrel\rightarrow}

\def\Re {\mbox{Re}}
\def\Im {\mbox{Im}}
\def\be{\begin{equation}}       \def\ee{\end{equation}}
\def\bea{\begin{eqnarray}}      \def\eea{\end{eqnarray}}
\def\bes{\begin{subequations}}  \def\ees{\end{subequations}}

\newcommand{\ci}[0]{\ensuremath{i}}
\newcommand{\e}[1]{\ensuremath{e^{#1}}}

\newcommand\effectivehamiltonian[1][t]{%
\begin{equation}
\label{eq:H_eff_triangular}
\begin{split}
\mathcal{H}_{\text{eff, th. lim.}}^{\mathcal{O}(5)} &=
N_s\cdot \epsilon_0 + J \sum_{\includegraphics[width=0.045\linewidth, trim={3cm 27cm 16cm 1.9cm},clip]{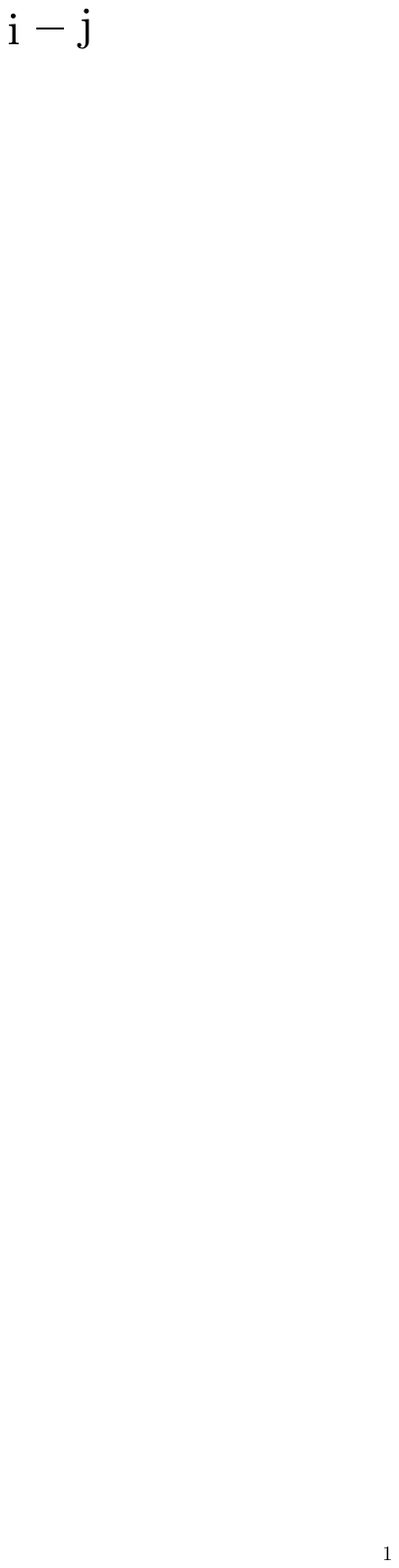}}
P_{ij}
+ \sum_{\includegraphics[width=0.045\linewidth, trim={3cm 26cm 16cm 1.9cm},clip]{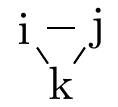}}
\left(K P_{ijk} + \text{h.c.} \right)
+ L^{\text{2sp}}_{\text{s}}\sum_{\includegraphics[width=0.063\linewidth, trim={3cm 27cm 15cm 1.9cm},clip]{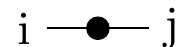}} P_{ij}
+ L^{\text{2sp}}_{\text{d}}\sum_{\includegraphics[width=0.058\linewidth, trim={3cm 26cm 15.5cm 1.9cm},clip]{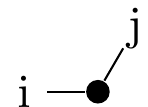}} P_{ij}\\
& + L^{\text{3sp}}_{\text{s}} \sum_{\includegraphics[width=0.075\linewidth, trim={3cm 26cm 14.5cm 1.9cm},clip]{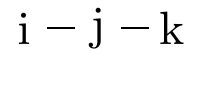}}
\left( P_{ijk} + \text{h.c.} \right)
+ \sum_{\includegraphics[width=0.055\linewidth, trim={3cm 25.5cm 15.5cm 2.1cm},clip]{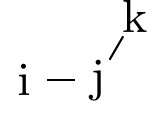}}
\left( L^{\text{3sp}}_{\text{d}} P_{ijk} + \text{h.c.} \right)
+ \sum_{\includegraphics[width=0.055\linewidth, trim={3cm 26cm 15.5cm 1.9cm},clip]{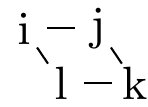}}
\left( L^{\text{4sp}}_{\text{r}}  P_{ijkl} + \text{h.c.} \right)
\\
& + \sum_{\includegraphics[width=0.065\linewidth, trim={3cm 26cm 15cm 1.9cm},clip]{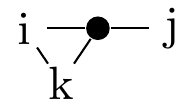}}
\left( L^{\text{3sp}}_{\text{cr}} P_{ijk} + \text{h.c.} \right)
+ \sum_{\includegraphics[width=0.065\linewidth, trim={3cm 26cm 14.8cm 1.9cm},clip]{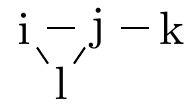}}
\left( L^{\text{4sp}}_{\text{cr}} \left( P_{lijk} + P_{kjli} \right) + \text{h.c.} \right)
+ L^{\text{4sp}}_{\text{pl}} \sum_{\includegraphics[width=0.055\linewidth, trim={3cm 26cm 15.5cm 1.9cm},clip]{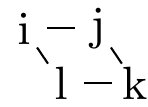}}
\left( P_{ikjl} + P_{kijl} + \text{h.c.} \right)
\\
& + L^{\text{4sp}}_{\text{Ka}} \sum_{\includegraphics[width=0.055\linewidth, trim={3cm 26cm 15.5cm 1.9cm},clip]{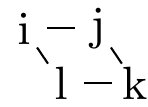}} P_{ij}P_{lk}
+ L^{\text{4sp}}_{\text{Kb}} \sum_{\includegraphics[width=0.065\linewidth, trim={3cm 26cm 14.8cm 1.9cm},clip]{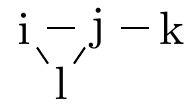}}
P_{il}P_{jk}
+ \sum_{\includegraphics[width=0.065\linewidth, trim={3cm 26cm 14.8cm 1.9cm},clip]{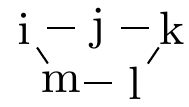}}
\left( L^{\text{5sp}}_{\text{r}} P_{ijklm} + \text{h.c.} \right)\ ,
\end{split}
\end{equation}
}

\begin{document}
\title{Time-reversal symmetry breaking Abelian chiral spin liquid in Mott phases of three-component fermions on the triangular lattice}

\author{C. Boos}
\affiliation{Institute for Theoretical Physics, FAU Erlangen-N\"urnberg, Germany}
\affiliation{Institute of  Physics, Ecole Polytechnique F\'{e}d\'{e}rale de Lausanne (EPFL), CH 1015 Lausanne, Switzerland}

\author{C.~J.~Ganahl}
\affiliation{Institut f\"ur Theoretische Physik, Universit\"at Innsbruck, A-6020 Innsbruck, Austria}

\author{M.~Lajk\'{o}}
\affiliation{Institute of Physics, Ecole Polytechnique F\'{e}d\'{e}rale de Lausanne (EPFL), CH 1015 Lausanne, Switzerland}

\author{P.~Nataf}
\affiliation{Univ. Grenoble Alpes, CEA INAC-PHELIQS, F-38000, Grenoble, France}

\author{A.~M.~L\"auchli}
\affiliation{Institut f\"ur Theoretische Physik, Universit\"at Innsbruck, A-6020 Innsbruck, Austria}

\author{K.~Penc}
\affiliation{Institute for Solid State Physics and Optics, Wigner Research Centre for Physics, H-1525 Budapest, P.O.B. 49, Hungary}

\author{K.~P.~Schmidt}
\affiliation{Institute for Theoretical Physics, FAU Erlangen-N\"urnberg, Germany}

\author{F.~Mila}
\affiliation{Institute of Physics, Ecole Polytechnique F\'{e}d\'{e}rale de Lausanne (EPFL), CH 1015 Lausanne, Switzerland}

\date{\today}

\begin{abstract}
We provide numerical evidence in favor of spontaneous chiral symmetry breaking 
and the concomitant appearance of an Abelian chiral spin liquid for three-component fermions on the triangular lattice described by an SU(3) symmetric Hubbard model with hopping amplitude $-t$ ($t>0$) and on-site interaction $U$. This chiral phase is stabilized in the Mott phase with one particle per site in the presence of a uniform $\pi$-flux per plaquette, and in the Mott phase with two particles per site without any flux.
Our approach relies on effective spin models derived in the strong-coupling limit in powers of $t/U$ for general SU$(N)$ and arbitrary uniform charge flux per plaquette, which are subsequently studied using exact diagonalizations and variational Monte Carlo simulations for $N=3$, as well as exact diagonalizations of the SU($3$) Hubbard model on small clusters. Up to third order in $t/U$, and for the time-reversal symmetric cases (flux $0$ or $\pi$), the low-energy description is given by the $J$-$K$ model with Heisenberg coupling $J$ and real ring exchange $K$. The phase diagram in the full $J$-$K$ parameter range contains, apart from three already known, magnetically long-range ordered phases, two previously unreported phases: i) a lattice nematic phase breaking the lattice rotation symmetry and ii) a spontaneous time-reversal and parity symmetry breaking Abelian chiral spin liquid. For the Hubbard model, an investigation that includes higher-order itinerancy effects supports the presence of a phase transition inside the insulating region, occurring at $(t/U)_{\rm c}\approx 0.07$ [$(U/t)_{\rm c} \approx 13$] between the three-sublattice magnetically ordered phase at small $t/U$ and this Abelian chiral spin liquid. 
\end{abstract}

\maketitle

\section{Introduction}

Quantum spin liquid phases are unconventional states of matter that have gained a lot of attention in the last decades due to their fascinating properties and possible future applications in quantum devices like quantum computers~\cite{Balents2010,Savary2016,Nayak2010}. From a theoretical point of view they are expected to emerge in strongly-correlated systems, for instance in Mott insulating phases. The recent progress in experiments with ultra cold atoms in optical lattices opens the exciting new possibility to simulate a broad variety of such quantum models~\cite{Bloch2008}. The optical lattice allows one to adjust the lattice type as well as the interaction strength, which can be tuned sufficiently to reach the Mott phase~\cite{Schneider2008,Taie2012,Hofrichter2016}. Furthermore, a synthetic static gauge field (the analog of magnetic flux in electronic systems) can be applied~\cite{Aidelsburger2013,Miyake2013}. The fundamental degrees of freedom can be adjusted by the choice of atom type. In particular fermionic alkaline earth atoms provide an SU($N$) symmetric spin degree of freedom with $N\leq 10$, because of an almost perfect decoupling between nuclear spin and electronic angular momentum~\cite{Cazalilla2009,Gorshkov2010,Scazza2014,Zhang2014,Cazalilla2014}. This new possibility to realize SU($N$) symmetric Hubbard models within the Mott phase~\cite{Taie2012,Hofrichter2016} creates a strong motivation for further theoretical investigations to identify potential realistic hosts of quantum spin liquid phases.

These systems can quite generally be described by an SU($N$) Hubbard model with a uniform "charge" flux defined by the Hamiltonian
\begin{equation}
 \mathcal{H} = - t \sum_{\langle i,j\rangle}\sum_{\alpha =1}^N(\e{i\phi_{ij}} c^{\phantom{,}\dagger}_{i\alpha} c^{\phantom{\dagger}}_{j\alpha}+\text{h.c.})+U\sum_{i,\alpha<\beta}n_{i\alpha} n_{i\beta}\ ,
\label{eq:ham_hubbard}
\end{equation}
where $t$ denotes the hopping amplitude of the fermions with color $\alpha = 1,\dots,N$. The Peierls phases $\phi_{ij}$ are chosen such that the flux per elementary plaquette of the lattice amounts to $\Phi$ and $U\ge0$ parametrizes the repulsive interaction strength. We use the convention $t\geq 0$, which is the natural sign for fermions hopping on a lattice with a quadratic spectrum at zero momentum. 

In the following, we will pay special attention to time-reversal symmetric models ($\Phi=0$ or $\pi$) since the main objective of this article is to identify chiral phases that spontaneously break time-reversal symmetry. As we shall see, in the Mott insulating phase with one particle per site, and with $t\geq 0$, such a phase is realized when $\Phi=\pi$, a situation to which we will refer as $\pi$-flux case. Now, on the triangular lattice, introducing a $\pi$ flux is equivalent to changing the sign of the hopping integral. Accordingly, for $N=3$ and in the Mott phase with two particles per site, such a phase is realized when $\Phi=0$ since
these two phases are related by a particle-hole transformation that changes the sign of the hopping term. 

In the strong-coupling limit $U \gg t$, and with an average filling of one particle per site, the system is in a Mott insulating phase, and effective magnetic descriptions are given in a basis of SU($N$) spins in the fundamental representation on every site.
Up to second-order in $t/U$ the effective description is given by the SU($N$) Heisenberg model
\begin{equation}
\mathcal{H} = J \sum_{\langle i,j\rangle} P_{ij}
\label{eq:Heisenberg}
\end{equation}
with the transposition operator $P_{ij}$, which exchanges the states on sites $i$ and $j$
\begin{equation}
P_{ij} = \sum\limits_{\alpha,\beta} \ket{\alpha_i \beta_j} \bra{\beta_i \alpha_j}\ ,
\label{eq:permutationoperator}
\end{equation}
where $\alpha$ and $\beta$ run over $N$ different colors~\footnote{Instead of the term 'colors' also 'flavors' is common in the literature}, i.e.~SU($N$) spins in the fundamental representation.
Up to second order, the coupling constant is given by $J=2t^2/U$. In a particular large $N$ limit of this model SU($N$) chiral spin liquids (CSL) were first reported~\cite{Hermele2009,Hermele2011}. In these phases parity and time-reversal symmetry are spontaneously broken, but their product is not~\cite{Hermele2009}, and most importantly, they exhibit topological order if the spectrum is gapped.
This shows up as a $2N$-fold degeneracy of the ground state on a torus~\cite{Hermele2011}.

By contrast, for SU(2), the Heisenberg model Eq.~\eqref{eq:Heisenberg} realizes a 120$^\circ$ long-range ordered state~\cite{Sachdev1992,Bernu1992,Capriotti1999,Zheng2006,White2007}. However, subleading four-spin interactions arising in order four in $t/U$ are known to trigger a first-order phase transition within the Mott phase to an exotic quantum disordered phase. Several investigations have led to numerical evidence in favor of a gapless spinon Fermi surface phase~\cite{Motrunich2005,Sheng2009,Yang2010,Kaneko2014}. However, recent numerical studies~\cite{Zhu2015,Hu2015,Hu2016,Iqbal2016,Wietek2017,Szasz2018,Hu2019} argue in favor of the presence of a chiral spin liquid or the existence of a gapped $\mathbb{Z}_2$ spin liquid or even a gapless Dirac spin liquid. So the definite identification of this exotic quantum phase within the $N=2$ Mott phase is not yet settled.

For SU(3) symmetric fermions on the triangular lattice, which is the system of interest in this work, it is useful to consider the expansion to third order in $t/U$. The effective model is then given by the $J$-$K$ model defined by the Hamiltonian
\begin{equation}
\mathcal{H} = J \sum_{\langle i,j\rangle} P_{ij} +  \sum_{\langle i,j,k \rangle}( K\, P_{ijk} + \text{h.c.})\ ,
\label{eq:ham_JK}
\end{equation}
where the first sum runs over all nearest-neighbor sites, and the second sum over all elementary triangles of the lattice. The operator $P_{ijk}=P_{ij}P_{jk}$ is a ring exchange operator that cyclically permutes the states between the sites $i$, $j$, and $k$. The coupling constants are given by \mbox{$J=2t^2/U-12\cos \left(\Phi\right) t^3/U^2$} and $K=-6 \e{i\Phi}t^3/U^2$, where $\Phi$ is the flux per triangular plaquette. 
A number of studies were performed for this $J$-$K$ model already~\cite{Bieri2012,Lai2013a,Lai2013b,Nataf2016}.
For purely real ring exchange, besides the conventional three-sublattice magnetically long-range ordered (3-SL LRO) phase~\cite{Tsunetsugu2006,Lauchli2006,Bauer2012}, two phases have been predicted by variational Monte Carlo (VMC): a $d_x+id_y$ CSL with spontaneous time-reversal symmetry breaking followed by a spin nematic phase~\cite{Bieri2012}. However, the uniform $\pi/3$-flux and $2\pi/3$-flux CSL states were not considered. A later mean-field study found that those are favored energetically for $0.41 \lesssim K \lesssim 6.0$ and $K\gtrsim 6.0$, respectively~\cite{Lai2013a}. 
This competition between many phases calls for investigations beyond mean-field and VMC. So far, this was only done for the $J$-$K$ model with purely imaginary ring exchange $K$, hence explicitly broken time-reversal symmetry, and the presence of robust CSL phases was confirmed for all $N$ from 3 to 9 by exact diagonalizations (EDs) on top of VMC~\cite{Nataf2016}. The first goal of the present paper is to extend this investigation to the case of purely real ring exchange, where we find in particular a previously unnoticed phase that breaks lattice-rotational symmetry, and a spontaneous time-reversal symmetry breaking CSL.

With respect to experiments, the most relevant open question is then whether the CSL phase is also present in the SU(3) Hubbard model, for which the $J$-$K$ model only provides a reliable description at very strong coupling $U$. Since EDs of the SU(3) Hubbard model are limited to very small system sizes (12 sites, see Sec.~\ref{sec:hubbardcomparison}), a direct investigation of the Hubbard model is difficult. One way to overcome this difficulty is to push the expansion in $t/U$ to higher order to extend the range of validity of the effective spin model. Indeed, not only is the effective model better because it contains more terms, but one can also test the effect of each additional order in the expansion and get a more precise idea of the range of validity of the effective model. So, the second goal of this paper is to push the expansion in $t/U$ to higher order (we will reach order five), and to identify the physics of the Hubbard model in the range where $U/t$ is not too large so that terms beyond nearest-neighbor exchange play a role, but still large enough to allow one to draw conclusions on the basis of the fifth-order expansion. As we shall see, ED complemented by VMC provides strong evidence in favor of a uniform $\pi/3$-flux CSL for moderate $K/J$, as well as for the Hubbard model in the Mott phase below $(U/t)_c \approx 13$.

The paper is organized as follows: At first we summarize the key results in Sec.~\ref{sec:keyresults}. In Sec.~\ref{sec:methods}, we review the various methods used in this manuscript - EDs (\ref{sec:methodsED}), variational Monte Carlo (\ref{sec:methodsVMC}), and the derivation of effective models using degenerate perturbation theory (\ref{sec:methodsderivationeffmodels}). Section~\ref{sec:JKmodel} is devoted to the $J$-$K$ model. We derive its phase diagram on the  basis of ED and VMC, and provide evidence for a CSL.
We then turn to the Hubbard model in Sec.~\ref{sec_hubbard}. We start with a detailed presentation of the fifth-order effective model in Sec.~\ref{sec_hubbard_effmodels}. We then benchmark this model in Sec.~\ref{sec:hubbardcomparison} by comparing its properties with those of the $\pi$-flux Hubbard model on 12 sites. Finally, we use this effective model to discuss the phase diagram of the $\pi$-flux Hubbard model and provide evidence for a CSL with spontaneous symmetry breaking in Sec.~\ref{sec:hubbardCSL}. The work is wrapped up in Sec.~\ref{sec:outlook} including a proposal for the experimental realization of the CSL.

\section{Key results}
\label{sec:keyresults}
\begin{figure}[t]
\begin{center}
\includegraphics[width=\linewidth]{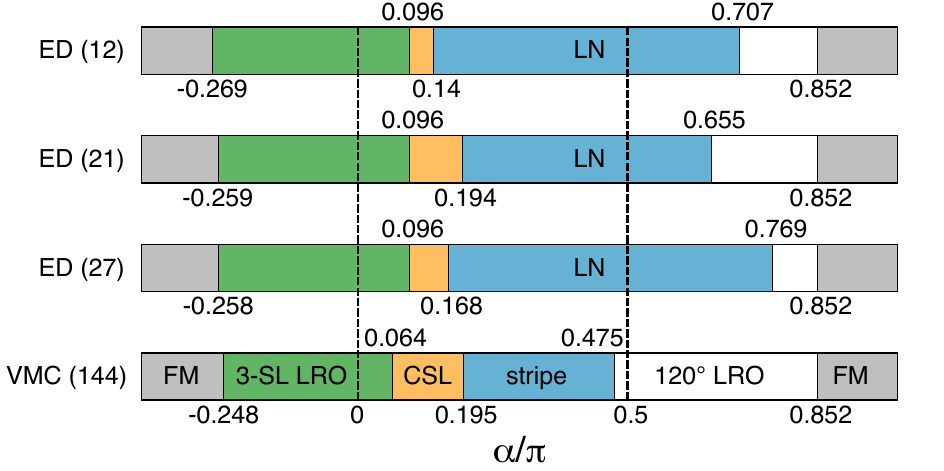}
\caption{Comparison of the predicted phase diagram for the triangular lattice $J$-$K$ model on the 12-, 21- and 27-site clusters from ED (top three rows) and the 144-site cluster from VMC (bottom row).
Around the Heisenberg point ($\alpha = 0$) the 3-sublattice ordered phase (3-SL LRO,green) is present. For increasing values of $\alpha$ up to $\alpha=\pi$ a $\pi/3$-flux chiral spin liquid (CSL,orange), a lattice nematic phase (LN,blue), a 120$^\circ$ LRO phase (white) and a ferromagnetically ordered phase (FM,gray) occur. The stripe state from VMC is expected to be closely related to the LN phase from ED as discussed at the end of Sec.~\ref{sec:JKmodel}.
}
\label{fig:phasediagramJKmodel}
\end{center}
\end{figure}

Before we give a detailed description of the methods and the full presentation of the results, we summarize our main findings.
Let us start with the ground-state phase diagram of the $J$-$K$ model with real ring exchange $K$ shown in Fig.~\ref{fig:phasediagramJKmodel}.
We use the $J$-$K$ notation, as well as the angle $\alpha$ in the range $-0.3 \leq \alpha / \pi \leq 1$, which relates to the previous parametrization by \mbox{$J= \cos \alpha$}, \mbox{$K= \sin \alpha$}~\cite{Bieri2012} . 
At the Heisenberg point, i.e.~$\alpha = 0$ ($K=0$, $J=1$), the well known 3-SL LRO phase is present. It is stable to the addition of a non-vanishing ring exchange of either sign.
For $\alpha \lesssim -0.25 \pi$ ($K<0$, $K/J \lesssim -1$) it eventually gets replaced by a ferromagnetic phase (FM).
For $K/J>0$ an increase of the ring exchange $K$ triggers a phase transition to a spontaneously time-reversal symmetry breaking $\pi/3$-CSL phase at $\alpha_{\rm c} \approx 0.096 \pi$ [$(K/J)_{\rm c} \approx 0.31$] in ED and $\alpha_{\rm c} \approx 0.064 \pi$ [$(K/J)_{\rm c} \approx 0.204$] in VMC.
This phase extends up to couplings $\alpha_{\rm c}\approx 0.19\pi$ [$(K/J)_{\rm c} \approx 0.67$], where the specific value depends strongly on the symmetry of the cluster under investigation.
Then, an unexpected phase with broken lattice-rotational symmetry occurs. This phase has not been reported in a previous VMC study~\cite{Bieri2012} and we call it a lattice nematic (LN) phase in the following
\footnote{A similar lattice nematic phase has been found in a bilinear-biquadratic $S=1$ spin model on the square lattice~\cite{Niesen2017}. Whether this phase extends up to the SU(3) point is however an
unsettled issue~\cite{Toth2010,Toth2012,Bauer2012,Hu2019b,Hu2020}.}.
In ED this LN phase extends up to $\alpha_{\rm c} \approx 0.7 \pi$ [$K>0, (K/J)_{\rm c} \approx -1.4$] for the sizes considered, i.e.~one requires negative $J$ to leave the LN phase.
In VMC however, the phase transition seems to occur at a smaller value of $\alpha_{\rm c} \approx 0.475 \pi$ [$K>0, (K/J)_{\rm c} \approx 13$].
We attribute this difference to both the strong finite size effects in ED as well as the possible necessary improvements in the VMC ansatz for the LN phase.
For larger $\alpha$ another phase is present.
This phase was previously introduced as a 120-nematic or $\mathcal{J}$-nematic phase in Ref.~\onlinecite{Bieri2012}. However, in the context of an SU(3) symmetric model 
we think it is best described by a 120$^\circ$ color ordered (120$^\circ$ LRO) phase, as it is essentially an SU(2) Heisenberg antiferromagnet embedded into an SU(3) symmetric system.
Finally, the FM occurs for $\alpha > 0.852 \pi$ [$K>0, (K/J) < -0.502$].

The most interesting aspect of the phase diagram is the presence of the spontaneous time-reversal symmetry broken CSL phase at intermediate coupling ratios $K/J$. We show that this phase has strong chirality correlations within a manifold of six low-lying eigenstates from the singlet sector on a torus in ED, reflecting the topological ground-state degeneracy.
The modular matrices of the apparent VMC states yield Abelian anyonic exchange statistics for four quasi-particles with topological spin $\pm 2\pi/3$ and chiral central charge $c=2$, in agreement with predictions for the $\pi/3$-flux CSL~\cite{Hermele2009}.

In view of the numerous phases discussed so far in the context of the $J$-$K$ model, and for the benefit of the reader, let us summarize the status of the various phases suggested so far:
\begin{itemize}
\item 3-sublattice LRO: present in all investigations so far.
\item $\pi/3$-flux CSL phase: predicted in Refs.~\cite{Hermele2009,Lai2013b}, not found in Ref.~\cite{Bieri2012}, confirmed here.
\item Lattice nematic: first mentioned here.
\item $2\pi/3$-flux CSL phase: predicted in Ref.~\cite{Lai2013b}, not confirmed here.
\item $d+id$: reported in Ref.~\cite{Bieri2012}, not confirmed here.
\item 120-nematic phase: first reported in Ref.~\cite{Bieri2012}, reinterpreted here as a 120$^\circ$ color ordered phase.
\item Ferromagnetic phase: present in all investigations that looked at this parameter range.
\end{itemize}

\begin{figure}[t]
 \begin{center}
\includegraphics[width=0.29\linewidth, angle = 270, trim={0.2cm 0.6cm 0.5cm 0.3cm},clip]{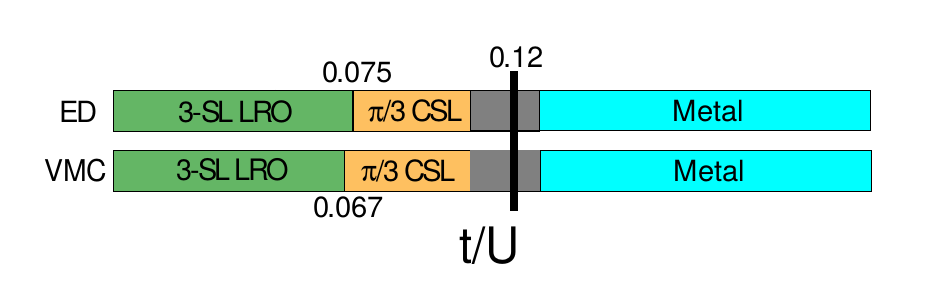}
  \caption{Phase diagram of the SU(3) Hubbard model with $\Phi=\pi$ on the triangular lattice. In the Mott-insulating phase for small $t/U$ the 3-SL LRO and $\pi/3$-CSL are found by ED and VMC. The metal-insulator transition is estimated in Sec.~\ref{sec:hubbardcomparison}
as $(t/U)_{\rm c}^{\text{mi}}\approx 1/8.5\approx 0.12$ (ED on 12 sites for the Hubbard model). The large uncertainty on this value and the nature of the Mott phase in that area is indicated by the grey area.}
  \label{fig:phasediagramHubbardmodel}
 \end{center}
\end{figure}

Second, we find that the physics of the $J$-$K$ model at small and intermediate coupling ratios, including the CSL, are present in the Mott phase with one particle per site of the SU($3$) $\pi$-flux Hubbard model. To this end, we derive the fifth-order effective spin model for the SU($N$) Hubbard model in the strong-coupling regime $t/U\rightarrow 0$ for general flux. For $\Phi = \pi$ corresponding to positive ratios $K/J$ and $N=3$, this spin model is qualitatively converged up to ratios $t/U$ including the phase transition point $(t/U)_{\rm c} \approx 0.075$ from ED and $(t/U)_{\rm c} \approx 0.067$ from VMC between the 3-SL LRO and the spontaneous time-reversal symmetry broken CSL phase.
The 12 site ED results for the Hubbard model suggest a metal-insulator transition at $(t/U)^{\text{mi}}_{\rm c}\approx 1/8.5\approx 0.12$. This implies the existence of the spontaneous time-reversal symmetry broken CSL in the Mott phase of the SU(3) Hubbard model with $\Phi=\pi$ on the triangular lattice as illustrated in Fig.~\ref{fig:phasediagramHubbardmodel}.

Finally, we note that the CSL is also present in the Mott phase with two particles per site of the SU($3$) Hubbard model without a flux. 

%
\section{Methods}
\label{sec:methods}
%
%

\subsection{Exact diagonalization}
\label{sec:methodsED}
%
%
\subsubsection{Spin models}
For the ED of spin models we employed a basis within irreducible representations (irreps) of the SU(3) group, which was introduced by two of the present authors~\cite{Nataf2014}.
Details are given in Appendix~\ref{sec:detailsonED}.
The method takes advantage of the full SU($N$) symmetry, and thus reduces the size of the Hilbert space under study much more effectively for $N>2$ than the usual approach of $U(1)$ color conservation together with lattice symmetries.
A drawback is that the approach is more contrived to obtain common quantum numbers like the lattice linear momentum or the angular momentum,
which are useful for identifing the appearing states~\cite{Wietek2017b}. 
To this end, we mainly used the following observables.
For magnetically ordered phases, the most prominent observable is the spin-spin correlator $S(i,j)$
between sites $i$ and $j$.
It is given by the expectation value of the transposition operator
\begin{equation}
 S(i,j) = \frac{\langle P_{ij}\rangle}{2}-\frac{1}{2N}\ .
\end{equation}
To find the ordering momentum, we studied the spin structure factor
\begin{equation}
 S(\vec{k}) = \sum\limits_{j=1}^{N_s-1} \e{-i\vec{k}(\vec{r}_j-\vec{r}_0)} S(j,0) + \frac{N^2-1}{2N}\ ,
\end{equation}
where $N_s$ is the number of sites and the last term gives the auto-correlator.
%
In addition, we also used the dimer correlator, which is the connected correlator of two bonds.
In terms of transposition operators, it reads
\begin{equation}
 D([i,j],[k,l]) = \langle P_{ij}P_{kl}\rangle-\langle P_{ij}\rangle\langle P_{kl}\rangle\ .
\end{equation}
The identification of the CSL phase is supported by the real space connected scalar chirality correlator $C(\langle i,j,k \rangle ,\langle l,m,n \rangle)$ for two individual triangles with sites $\langle i,j,k\rangle$ and $\langle l,m,n\rangle$ oriented in the same way.
The scalar chirality for a single triangle is defined as
\begin{equation}
 \langle\chi(i,j,k)\rangle = \frac{\mathrm{i}}{4}\langle P_{ijk}-P_{kji}\rangle\ .
\end{equation}
Then, the connected scalar chirality correlator is given by
\begin{equation}
\begin{split}
 C(\langle i,j,k \rangle ,\langle l,m,n \rangle) = \langle \chi(i,j,k)\;\chi(l,m,n)\rangle \\
 - \langle\chi(i,j,k)\rangle \langle\chi(l,m,n)\rangle\ ,
\end{split}
\end{equation}
With the overall chirality or chirality signal we refer to the sum of all connected chirality correlators where
$\langle i,j,k\rangle$ is a fixed reference triangle and $\langle l,m,n\rangle$ are all triangles which share no site with the reference triangle and are labeled in the same orientation. 
For a CSL, which features long-range chiral correlations, the chirality signal should be strong and the real space connected chirality correlator should be positive and 
uniform across the lattice.

The calculations were performed on the 12-site, 21-site, and 27-site clusters shown in Fig.~\ref{fig:edclusters}.

\subsubsection{Hubbard Model}
\label{sssec:methodsEDHubbard}
The full Hilbert space of the SU(3) Hubbard model contains states with multiple occupied sites and
we cannot employ the same basis as before.
Instead, we use an ordinary U($1$) quantum number basis, where we label sectors by the numbers of fermions of each of the three colors $a, b, c$ as $(N_a, N_b, N_c)$.

We also attempt to pin down the metal-insulator transition point by computing the partice-hole charge gap.
For a general SU($N$) Hubbard model this can be done by calculating the lowest energies of three different Hilbert space sectors, namely the sector with the same number of particles per color at $1/N$ filling, 
the sector with one additional fermion, and the sector with one fermion less for one color compared to the nominal filling.
Then, the particle-hole charge gap is
\begin{equation}
\Delta_\mathrm{charge} = E_0(+1)-2\cdot E_0(0)+E_0(-1)\ ,
\label{eq:Deltacharge}
\end{equation}
where the numbers in parenthesis label the deviation from the Mott filling.
In order to obtain an estimate for the metal-insulator transition, one has to extrapolate the large-$U$ linear part of the particle-hole charge gap.
The point where the extrapolated line crosses zero, so where the charge gap closes, is an estimate for the location of the metal-insulator transition.

\subsection{Variational Monte Carlo}
\label{sec:methodsVMC}
The VMC approach is based on projected $1/3$ filled non-interacting parton wave functions. We start from a nearest-neighbor tight-binding model
\begin{equation}
\begin{split}
\mathcal{H}_\alpha =\sum_\alpha  \Big[ -\sum_{\left\langle i,j \right\rangle} \left( t_{i,j}^\alpha  f_{i,\alpha} ^\dagger f_{j,\alpha}^{\phantom{\dagger}} +\text{h.c} \right)  -  \sum_i h_{i,\alpha}^{\phantom{\dagger}}  f_{i,\alpha} ^\dagger f_{i,\alpha}^{\phantom{\dagger}} \Big]\ ,
\end{split}
\label{eq:tightbinding}
\end{equation}
where $\alpha$ denotes the color, and  $ f_{i,\alpha} ^\dagger (f_{i,\alpha}^{\phantom{\dagger}})$ are  creation (annihilation) operators of an $\alpha$-fermion (parton) at site $i$. 
For each color the one-fermion states are filled up to one-third filling.
Then, a Gutzwiller projection~\cite{Yokoyama1987,Gros1989,Corboz2012} excludes the charge fluctuations.
The energies, symmetries, and overlaps of these variational states can be evaluated using a Monte Carlo sampling.
The hopping amplitudes and the on-site fields of the tight-binding model serve as variational parameters. 
Similar studies of the \mbox{$J$-$K$ model} for real $K$ were carried out by Bieri et al.~\cite{Bieri2012}.
They only considered time-reversal symmetry breaking through complex pairing terms ($\text{p}+i \text{p}$ and $\text{d} + i \text{d}$ states in their paper) and did not consider any state with complex hoppings and non-time-reversal invariant flux configurations.
The latter is a natural way to create CSL variational states~\cite{Hermele2009, Lai2013b, Nataf2016}.
In the following, we provide a list of the states we considered.
We carried out these simulations on two clusters of $6\times6=36$ and $12\times12=144$ sites, both compatible with all the listed scenarios.
A detailed description of our results is given in Sec.~\ref{sec:VMCresults}, where we also compare the energies of the most competitive states to the ones proposed in Ref.~\onlinecite{Bieri2012}.

\emph{CSL states---} 
To create spin liquid variational states, we set all the $h_{i,\alpha}$ on-site terms to zero, and the $t_{i,j}^\alpha$ hopping amplitudes to unity. The phases of the hopping amplitudes are such that the
flux on every triangular plaquette is the same. We considered the cases with $m\pi/6$-flux $(m \in \mathbb{Z})$ per plaquette. We also made calculations on states with opposite fluxes on up and down triangles with fluxes $\pm m \pi/3$.
 
\begin{figure}[t]
\begin{center}
\includegraphics[width=\linewidth]{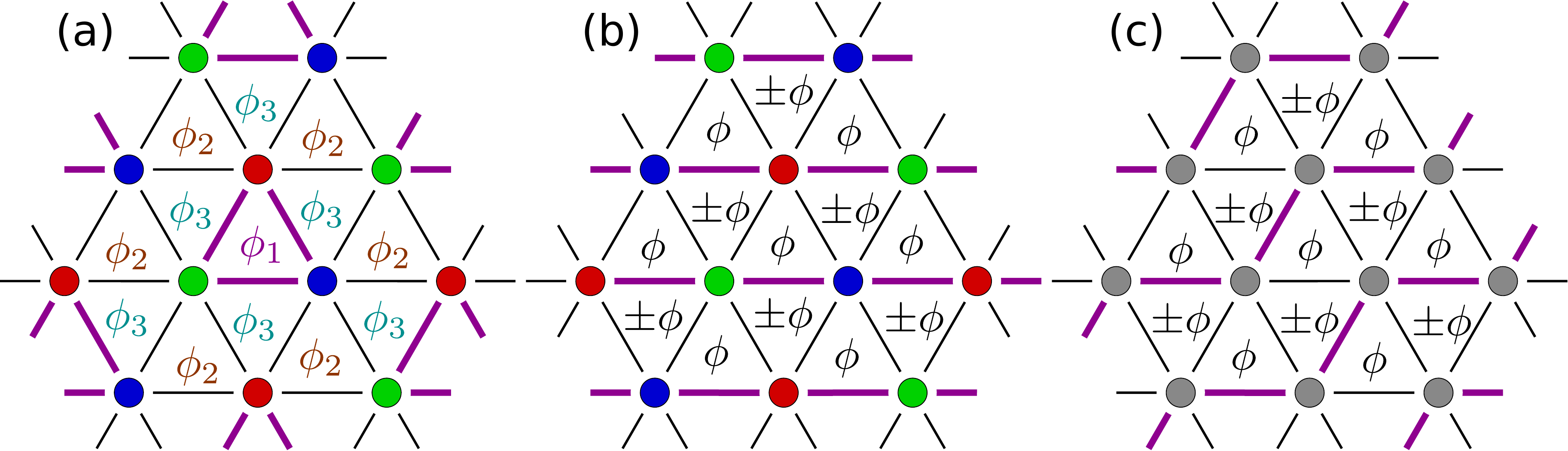}
\caption{Illustration of the considered hopping configurations in the VMC calculations.
(a) For the all-up plaquette order the magnitude of the hopping amplitudes on the thick purple bonds was changed between 0.5 and 2. The phases $\phi_1$, $\phi_2$, and $\phi_3$ were either chosen to be uniform with values $m\pi/6$ ($m \in \mathbb{Z}$), or set to a combination of $0$ or $\pi$.
For the 3-SL LRO (shown by the color of the vertices) the on-site chemical potential was changed from $0$ to $2$.
For the (b) stripe order, and (c) zigzag stripe order, the strengths of the hopping on the purple bonds are set to 1, and the strengths on the thin black bonds between the stripes are tuned between 0 and 2.
In both
stripe cases we considered either uniform flux configurations or opposite fluxes on up and down triangles with $\phi = \{0, \pi/3, \pi/2,  2\pi/3, \pi\}$.}
\label{fig:hoppingplot}
\end{center}
\end{figure}

\emph{Plaquette ordered states---} 
By increasing the magnitude of the hopping amplitudes around distinct plaquettes in Eq.~\eqref{eq:tightbinding}, we can create variational states with lower bond energies around these plaquettes. In particular, we focused on the possibility of strong bonds around triangular plaquettes corresponding to the formation of $\mbox{SU}(3)$ singlets. We considered all the possible plaquette coverings on the 36-site cluster, with a uniform zero- or $\pi$-flux per plaquette, changing the magnitude of hoppings around the plaquettes from 0.5 to 2 (the magnitude of the other hopping amplitudes were set to  1). We examined the all-up plaquette covering [shown in Fig.~\ref{fig:hoppingplot}(a)] in more detail for both the 36- and the 144-site system. On top of changing the magnitude of the bonds around the triangular plaquettes, we considered various flux configurations with either zero- or $\pi$-flux per plaquette for each of the three different types of triangles [marked with $\phi_1$, $\phi_2$, and $\phi_3$ in Fig.~\ref{fig:hoppingplot}(a)]. We also considered the combination of the all-up plaquette order with a uniform $m\pi/6$-flux $(m \in \mathbb{Z})$ per plaquette. 

\emph{Stripe order---}
Similarly to the plaquette ordering, by strengthening the hoppings along chains
in Eq.~\eqref{eq:tightbinding}, we
generate variational states of weakly coupled SU(3) chains. We considered the cases of straight and zigzag chains [c.f. Fig.~\ref{fig:hoppingplot}(b) and (c)] with \mbox{$0 \leq t_{\text{inter-chain}}/t_{\text{intra-chain}}\leq 2$} on the 36- and 144-site clusters. Note, that the zigzag chains are compatible with the 36- and 144-site clusters, but not with the 21- and 27-site clusters used for ED. On top of varying $t_{\text{inter-chain}}/t_{\text{intra-chain} }$, we considered uniform flux-configurations of $0$-, $\pi/3$-, $\pi/2$-,  $2\pi/3$-, and \mbox{$\pi$-flux}, as well as opposite fluxes on up and down triangles with the same values.
The straight stripe $\pi$-flux state is related to the LN phase from ED in Sec.~\ref{sec:EDresults}.

\emph{Color order---}
If the chemical potentials are different for different colors, the resulting variational states will have $\mbox{SU}(N)$ symmetry breaking color order. We considered states with a three-sublattice color order as shown in Fig.~\ref{fig:hoppingplot}(a). On each sublattice the on-site term for only one color is non-zero, i.e.~$ h_{i, \alpha}= h$ if $i$ is in sublattice $\alpha$. The magnitude of the on-site field is varied between $0<h<2$. We also combined this color order with the all-up plaquette order and the straight stripe order of the previous points.

Gutzwiller projected states can also be used to access different topological sectors by introducing twisted boundary conditions in Eq.~\eqref{eq:tightbinding} before Gutzwiller projection~\cite{Nataf2016}.
Then, if the number of non-zero eigenvalues of the overlap matrix is independent of how many different twisted boundary conditions are considered, it gives the number of linearly independent states.

\subsection{Derivation of effective models}
\label{sec:methodsderivationeffmodels} 
The effective model for the Mott phase of a general SU($N$) Hubbard model with an arbitrary flux for an average filling of one particle per site in the strong-coupling limit is stated and analyzed in Sec.~\ref{sec_hubbard_effmodels}.
%
%
Here, we briefly motivate the underlying calculations, which are based on a linked-cluster expansion (LCE) and degenerate perturbation theory.
A similar approach was successfully applied to the SU(2) Hubbard model on the triangular lattice in Ref.~\onlinecite{Yang2010}.

A LCE is a technique to extend results on small finite clusters to larger clusters or to the thermodynamic limit.
We use a very appealing LCE approach along the lines of a white-graph expansion~\cite{Coester2015}, which we employ to simplify the subtraction process, as well as to include complex phases. This is explained in detail in Appendix~\ref{sec:detailsonLCE}.

On every linked cluster, we use degenerate perturbation theory~\cite{Kato1950, Takahashi1977, MacDonald1988, MuellerHartmann2002, Yang2010}
%
about the strong-coupling limit $t\rightarrow 0$ in Eq.~\eqref{eq:ham_hubbard}, and determine effective descriptions in the subspace of states with exactly one fermion per site.
%
%
A link is created by the perturbative hopping of a fermion between two sites.
We find that in order $k$ a linked cluster only yields a non-vanishing contribution if the number of links that are part of a loop plus twice the number of links that are not part of a loop is smaller or equal to the order.
%
%
In order 5 on the triangular lattice, six linked-clusters give a non-zero contribution, namely the dimer, the trimer, the triangle, a triangle with one additional site, a four-site loop, and a five-site loop.
%
For each cluster one derives the associated reduced effective Hamiltonian,
which can be written in a compact form
%
\begin{equation}
\mathcal{H}_{\text{eff}}(t/U,\Phi)=\sum_{(i,j)} A_{ij} P_{ij} + \sum_{(i,j,k)} B_{ijk} P_{ijk} + ...\ ,
\label{eq:eff_ham}
\end{equation}
where the coupling constants depend on $t/U$ and $\Phi$.
Here and in the following, all effective Hamiltonians and effective couplings are given in units of $U$.
%
Every linked cluster yields a different set of exchanges that are embedded on the full system in the second step.
Some of these exchanges are unique to the specific cluster, whereas other interactions emerge from contributions of a variety of linked-clusters.
We perform these calculations for all interactions in the effective model in order 5 for the thermodynamic limit (Sec.~\ref{sec_hubbard_effmodels}) and in order 4 for the 12-site cluster with PBCs.

To improve and assess the quality of convergence of the coupling constants, we employ Pad\'e extrapolants~\cite{Guttmann1989}.
These are rational functions and can be denoted with $[m,n]$ where the degree of the numerator is $m$ and that of the denominator is $n$.
At certain parameters, unphysical divergences emerge, which have to be excluded.
%
%
%
\section{$J$-$K$ model}
\label{sec:JKmodel}
In this section, we determine the phase diagram of the $J$-$K$ spin model for arbitrary real values of the $J$ and $K$ parameters by ED and VMC.
The results of both methods are compared at the end of the section.

\subsection{Exact diagonalization}
\label{sec:EDresults}

%
%
\begin{figure}[t]
 \begin{center}
  \includegraphics[width=0.95\linewidth]{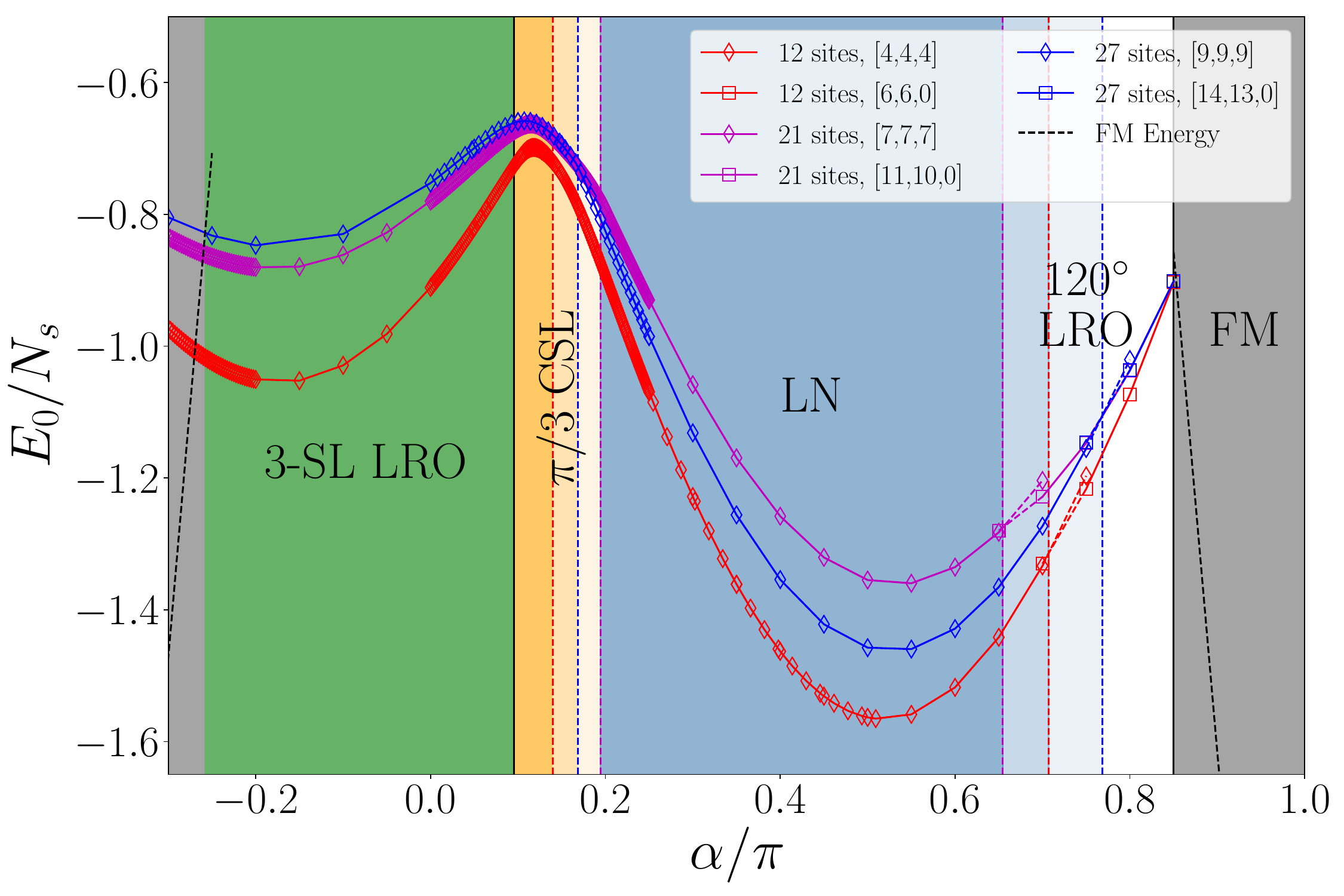}
\caption{Ground-state energies from ED on clusters with \mbox{$N_s=\{12, 21, 27\}$} sites for the $J$-$K$ model with the coupling constants $J= \cos \alpha$ and $K=\sin \alpha$.
Different point shapes indicate different symmetry sectors. The singlet sector $[N_s/3, N_s/3, N_s/3]$ yields the ground states of the 3-SL LRO, CSL and LN phase below $\alpha/\pi \approx 0.7$. For $0.7 \lesssim \alpha/\pi \lesssim 0.85$ the 120$^\circ$ color order state, which contains only two colors from the sector $[N_s/2,N_s/2,0]$ or $[(N_s+1)/2,(N_s-1)/2,0]$, is present. In the FM phase for $\alpha/\pi \gtrsim 0.85$, where the energy per site is identical for all clusters, only a single color is present and therefore the ground states lie in the symmetry sectors $[N_s, 0, 0]$. Wherever the phase boundaries are not the same for different lattice sites $N_s$, the color of the boundary indicates the corresponding cluster size.}
\label{fig:JKgroundstatespectrumED}
\end{center}
\end{figure}

%
%
\begin{figure}[b]
	\centering
		\includegraphics[width=0.95\linewidth]{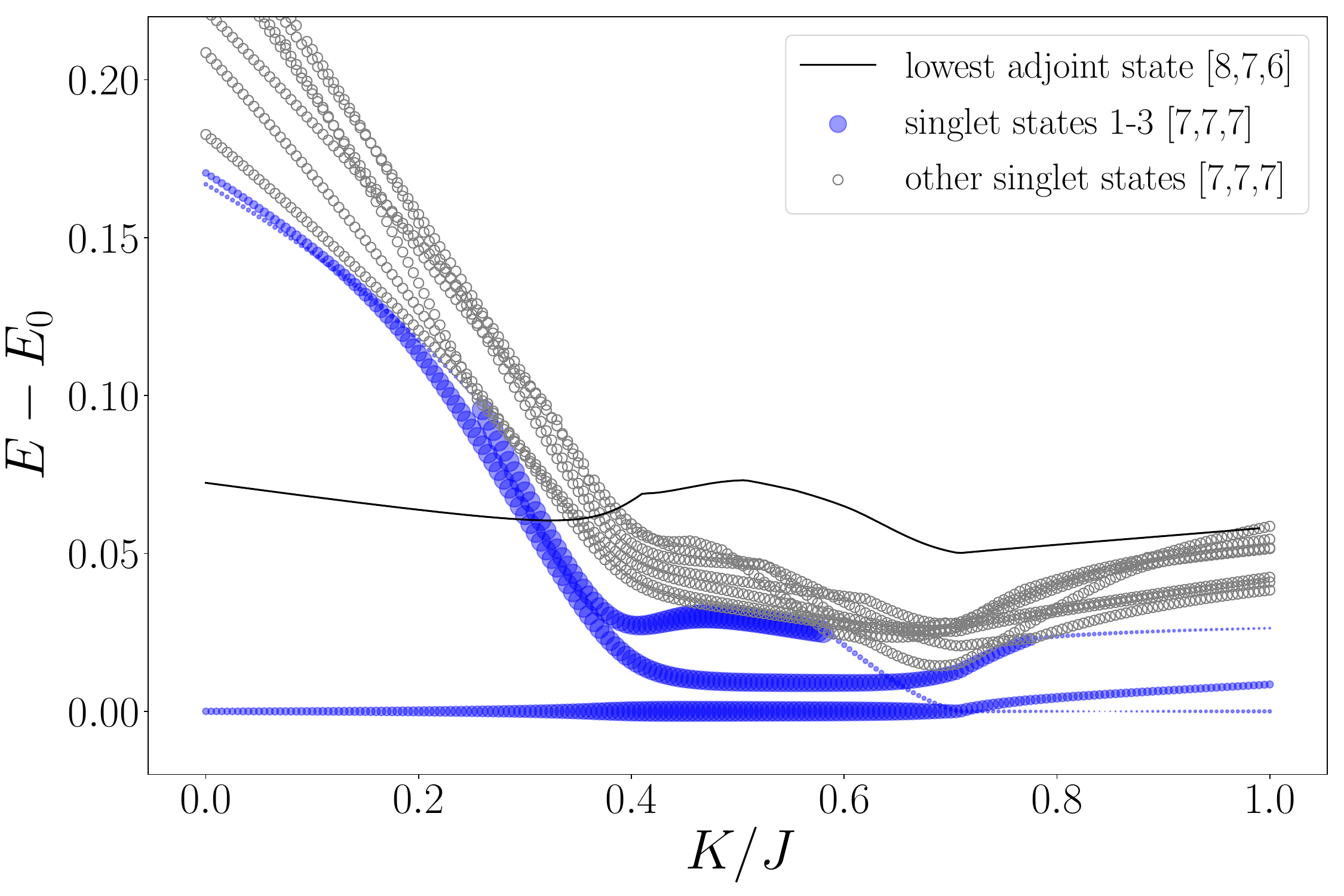}
	\caption{Spectrum of the $J$-$K$ model on the 21-site cluster from ED. The marker size corresponds to the overall chirality signal and is plotted for the lowest three states (the degeneracy of these states is  1, 4, and 1). The three (six when we count the degeneracies) low-lying singlets with strong chiral signal indicate the presence of a CSL phase in an extended parameter space at intermediate values of $K/J$.}
	\label{Fig:S_Spectra_chiral_states}
\end{figure}

%
%
\begin{figure}[t]
	\centering
		\includegraphics[width=0.95\linewidth]{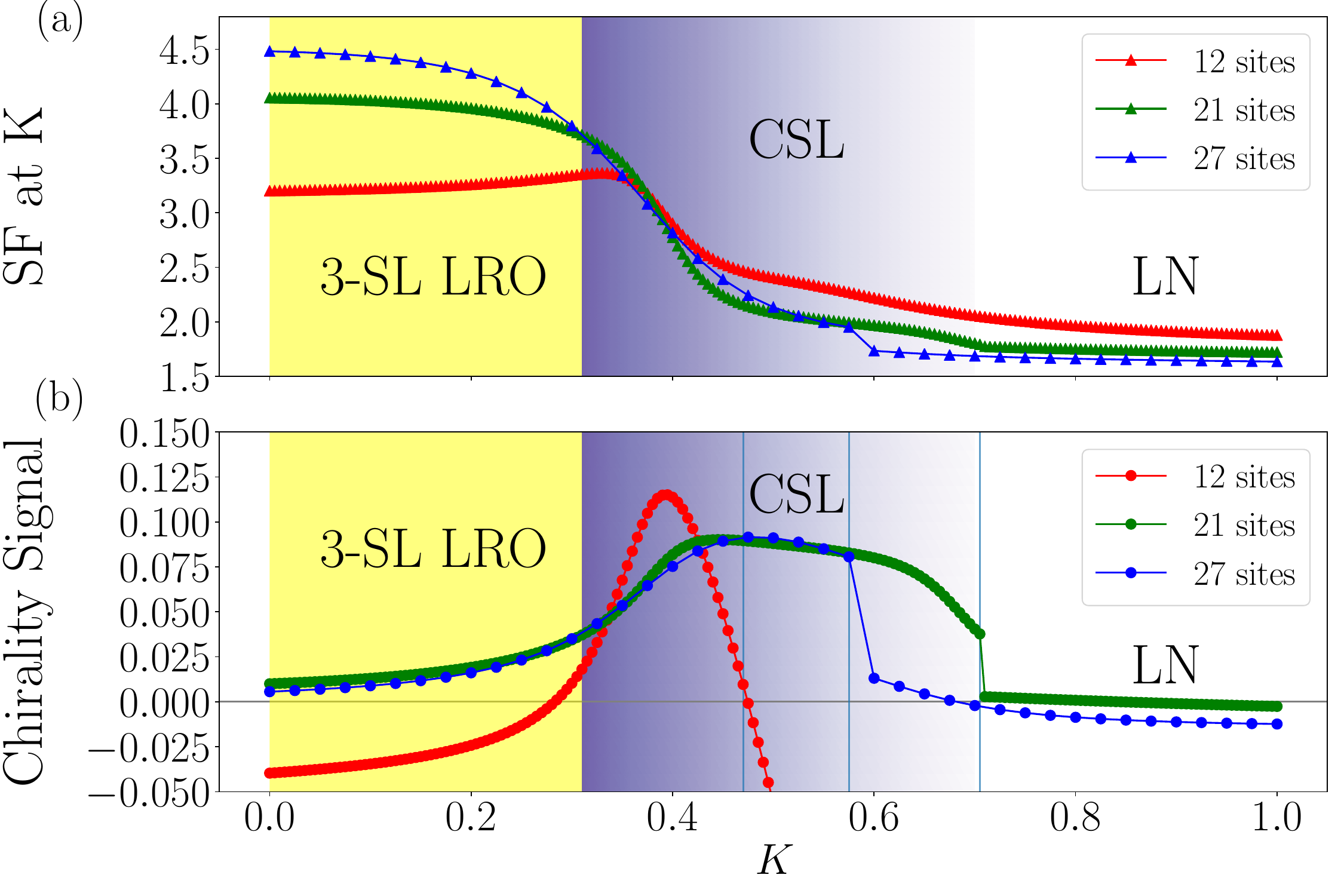}
	\caption{Structure factor at the K point (top) and chirality signal per lattice site (bottom) for the $J$-$K$ model from ED on the 12-, 21- and 27-site clusters. Coming from the 3-SL LRO phase at small ratios $K/J$, the structure factors at the ordering momentum $K$ decrease as the chirality signals increase, indicating the CSL. In the regime, where the chirality signal decreases, a LN phase occurs.}
	\label{Fig:S_Chirality_SF}
\end{figure}

%
%
%
\begin{figure}[b]
	\centering
		\includegraphics[width=0.95\linewidth]{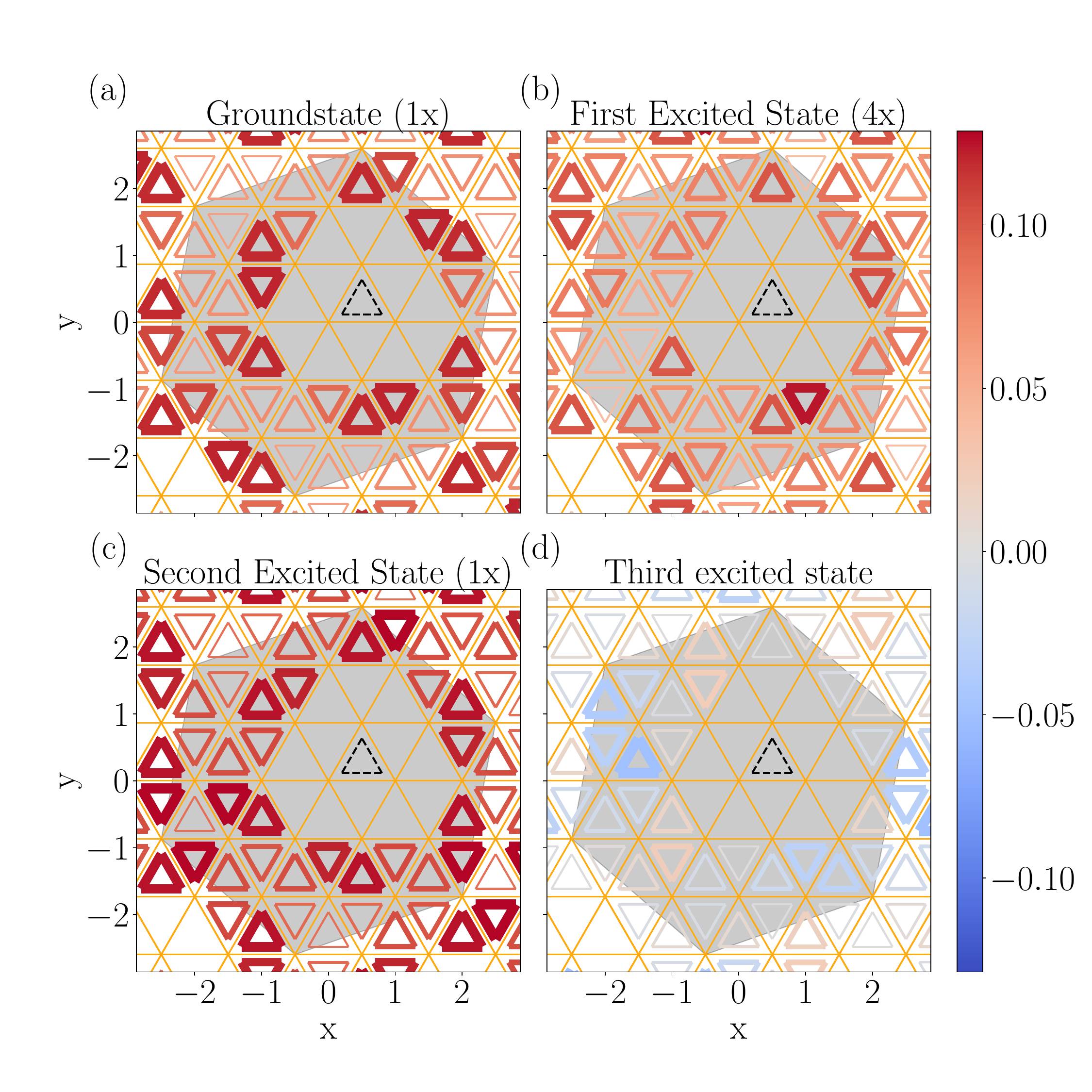}
	\caption{Real space connected chirality correlator for the $J$-$K$ model from ED on the 21-site cluster at $K/J=0.455$. The degeneracy of the first three singlet states is 1-4-1. While the lowest three singlet levels show long range chiral ordering, the fourth one does not.}
	\label{Fig:S_Realspace_Chirality}
\end{figure}

%
%
\begin{figure}[b]
	\centering
		\includegraphics[width=0.95\linewidth]{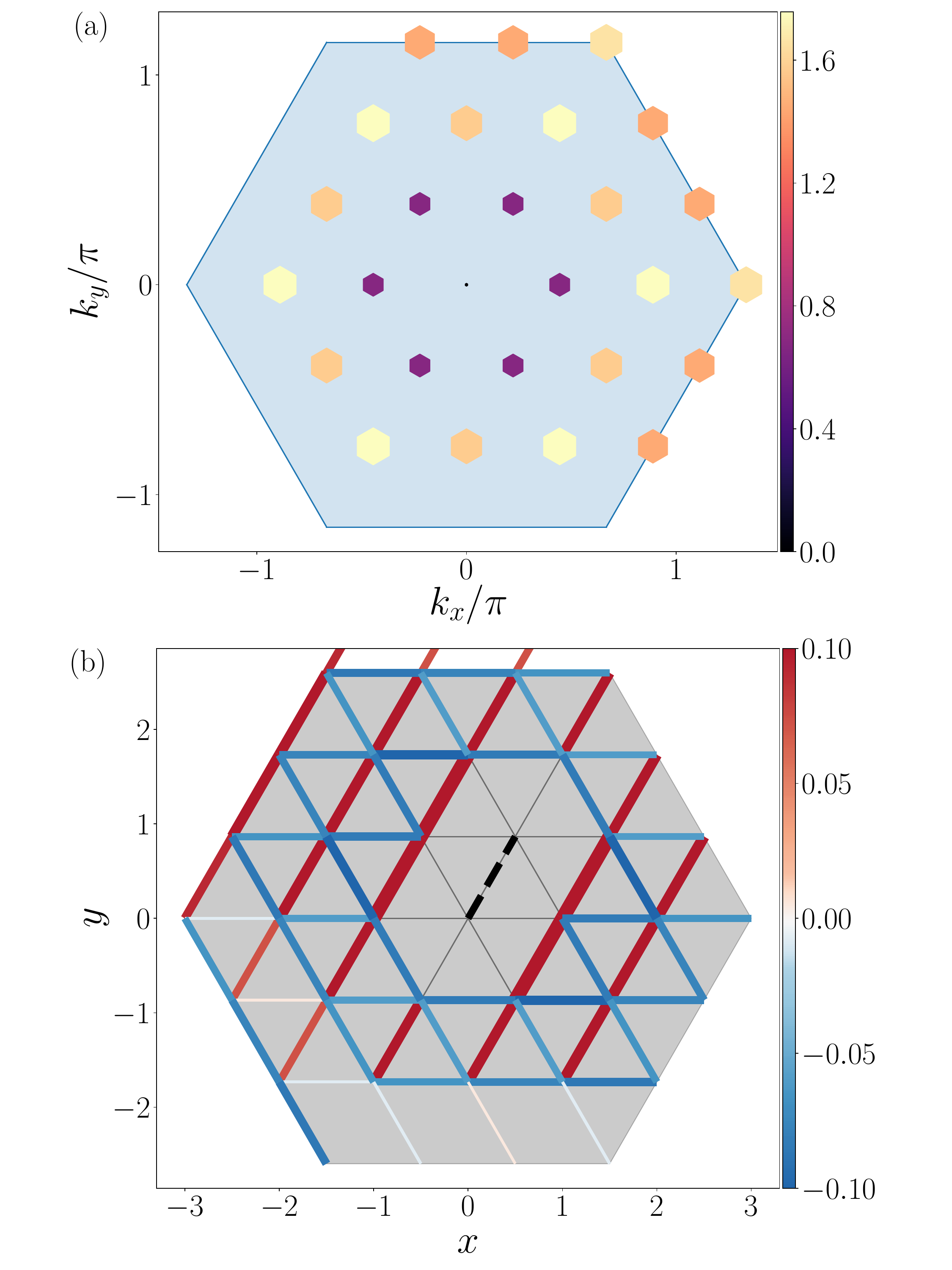}
	\caption{Spin structure factor (top) and dimer-dimer correlations (bottom) of the $J$-$K$ model in the LN phase from ED on the 27-site cluster. The structure factor peak is close to the X point. The dimer-dimer correlations clearly show that the rotational symmetry of the lattice is broken in this phase.}
	\label{Fig:S_SF_Dimer_nematic}
\end{figure}

Using ED, we studied the $J$-$K$ model on the $12$-, $21$-, and $27$-site triangular clusters with periodic boundary conditions (PBCs). 
We start with a discussion of the ground-state energy per site plotted as a function of the parameter $\alpha$ in
Fig.~\ref{fig:JKgroundstatespectrumED}. We show the energies for three different SU(3) irreps: the singlet, the ferromagnet and the irrep which corresponds to the lowest effective SU(2) sector (for reasons which become clear below). We discuss the evidence leading to the labeling of the different phases in the following. Note that the CSL seems to be squeezed at the interface between two quite extended phases, similar to scenarios in SU(2) quantum magnets, where spontaneous time-reversal symmetry breaking CSL have been observed~\cite{Gong2014,He2014,Wietek2015}.

The appearance of the CSL when increasing $K>0$ ($J>0$), or $\alpha>0$, can be understood with the low-energy spectrum of the 21-site cluster from ED in Fig.~\ref{Fig:S_Spectra_chiral_states}, where the marker size illustrates the overall chirality signal for the lowest {\em three} singlet energy levels (whose degeneracies are not resolved here). 
For small values of $K/J$ the first excited state above the singlet ground state is in the adjoint irrep $[8,7,6]$, which corresponds to the tower of states expected for the 3-SL LRO phase~\cite{PencLaeuchli2010}.
With an increasing ratio $K/J$ two singlets cross the adjoint irrep around $(K/J)_{\rm c} \approx 0.31$ ($\alpha_{\rm c} \approx 0.096 \pi$) and become the lowest excitations. In the same parameter range the total chirality of the ground state, as well as that of the two levels from the singlet sector above, increases, which is an indication for a chiral phase.
More importantly, the degeneracies of the three lowest singlet levels have been determined to be 1-4-1 corresponding to a total of six states, which may form a six-fold degenerate ground state in the thermodynamic limit.
Such a degeneracy is associated with spontaneous chiral symmetry breaking in the SU(3)-symmetric $\pi/3$-CSL on a torus~\cite{Hermele2011}. However, these states are not very well separated from higher excited states on the 21-site cluster. The signature is more pronounced than on the 12-site cluster though. On the 27-site cluster the splitting between the six low-energy states and the states above is comparable to that on the 21-site cluster.

In the upper panel of Fig.~\ref{Fig:S_Chirality_SF} the magnetic structure factor at the ordering momentum of the 3-SL LRO phase, the K point, is shown. It is extensive in the ordered phase but decreases as the chirality signal increases, which is displayed in the lower panel of Fig.~\ref{Fig:S_Chirality_SF}. This indicates a decay of magnetic ordering. The phase boundary in both panels of Fig.~\ref{Fig:S_Chirality_SF} is determined 
by the crossing of the two lowest singlet excitations with the adjoint irrep.
The connected scalar chirality correlator for $K/J=0.455$ ($\alpha \approx 0.14 \pi$), where the first and second excitations are both singlets, is illustrated in Fig.~\ref{Fig:S_Realspace_Chirality}. We find an almost uniform distribution in the connected chirality correlator for the first three singlets, as expected for a CSL phase. The third excited state shows no pronounced chiral order anymore.
The signature of the CSL phase varies in its extension in parameter space for different cluster sizes. It is larger for the 21-site cluster than for the 12- and 27-site clusters. This is due to the nature of the adjacent phase at larger values of $K/J$ ($\alpha$), which seems to have its peak in the magnetic structure factor close to the X point. While the 12 sites cluster has this particular point, the 21 and 27 sites cluster do not, which may lead to an increased parameter space for the CSL phase on those clusters.

%
%
\begin{figure}[b]
	\centering
		\includegraphics[width=1\columnwidth]{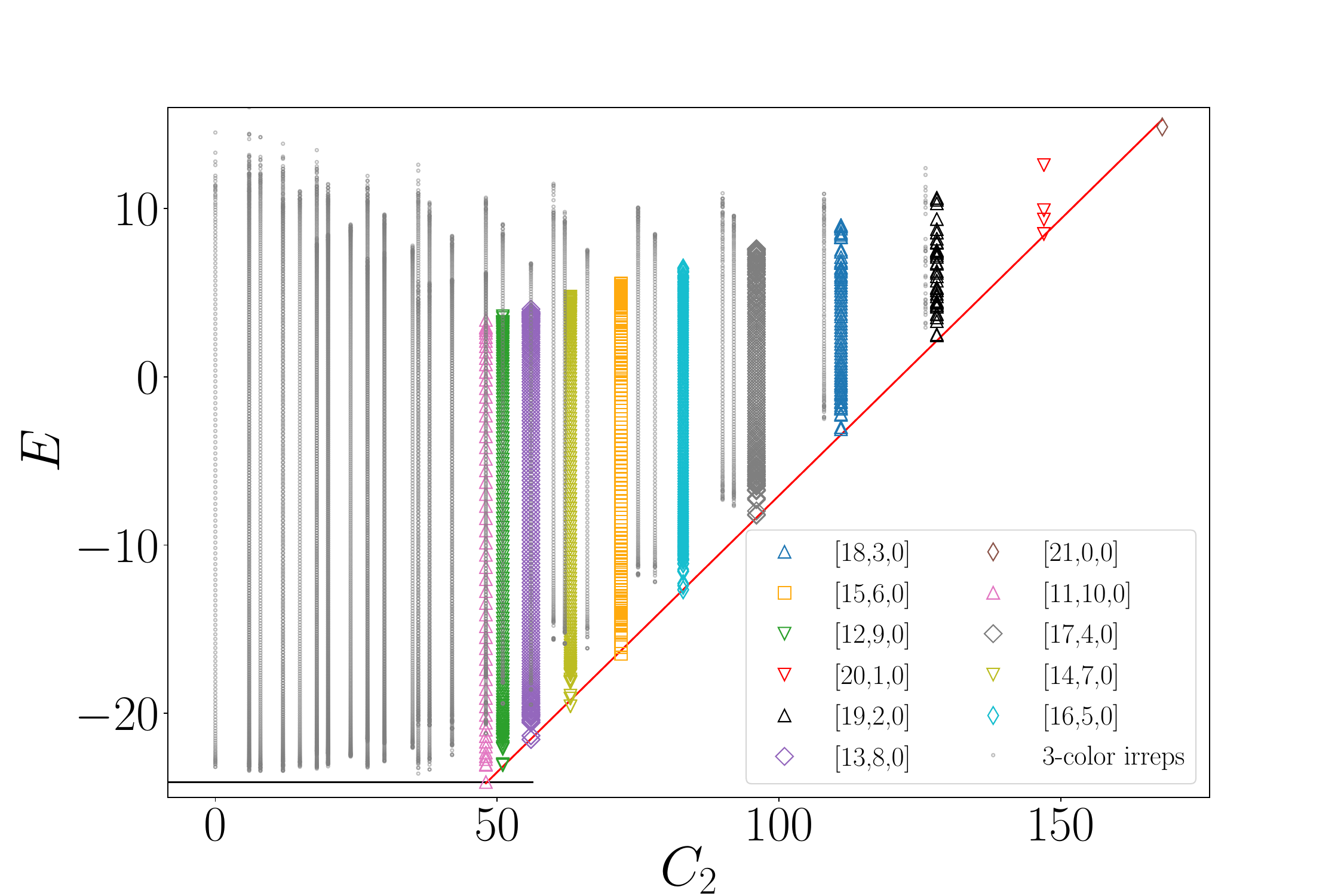}
	\caption{Spectrum of the SU(3) $J$-$K$ model at $\alpha=3\pi/4$ from ED on the 21-site cluster as a function of the quadratic Casimir $C_2$ of each irrep. The ground state lies in the $[11,10]$ irrep of SU(3), where only two of the three possible colors are present. The black line is a guide to the eye illustrating that the ground state is indeed not a singlet. The red line is a linear fit through the lowest eigenvalue in each two-color irrep corresponding to the effective SU($2$) tower of states in the spectrum. Further information on the quadratic Casimir operator is given in Appendix \ref{app:quadraticcasimir}.}
	\label{Fig:S_TOS_SU2}
\end{figure}

The phase next to the CSL is a previously unreported spatial symmetry breaking phase which is characterized by strong bonds along one lattice direction and weak bonds along the other lattice directions. It therefore breaks the rotational symmetry of the lattice, leaving a strong signature in the dimer-dimer correlations, as can be seen in Fig.~\ref{Fig:S_SF_Dimer_nematic}. Given the limitations in reachable cluster sizes using ED, it remains an open question whether there is any sort of magnetic ordering in this LN phase or not.

Increasing $K/J$ ($\alpha$) further leads to an SU($2$)-like behavior with one color less. We confirm this by finding an SU($2$) tower of states at large $K/J$ where the ground state is in the irrep $[N_s/2,N_s/2]$ (or $[(N_s-1)/2+1,(N_s-1)/2]$ if the number of lattice sites $N_s$ is odd) of SU(3) as shown in Fig.~\ref{Fig:S_TOS_SU2}. According to Eq.~\eqref{eq:reduction}, the ground-state energies in this region can be compared to ED results on the nearest-neighbor spin-1/2 triangular lattice~\cite{Bernu1994}, and we find a perfect match.

\subsection{Variational Monte Carlo}
\label{sec:VMCresults}
\begin{figure}[ht]
\begin{center}
\includegraphics[width=0.45\textwidth]{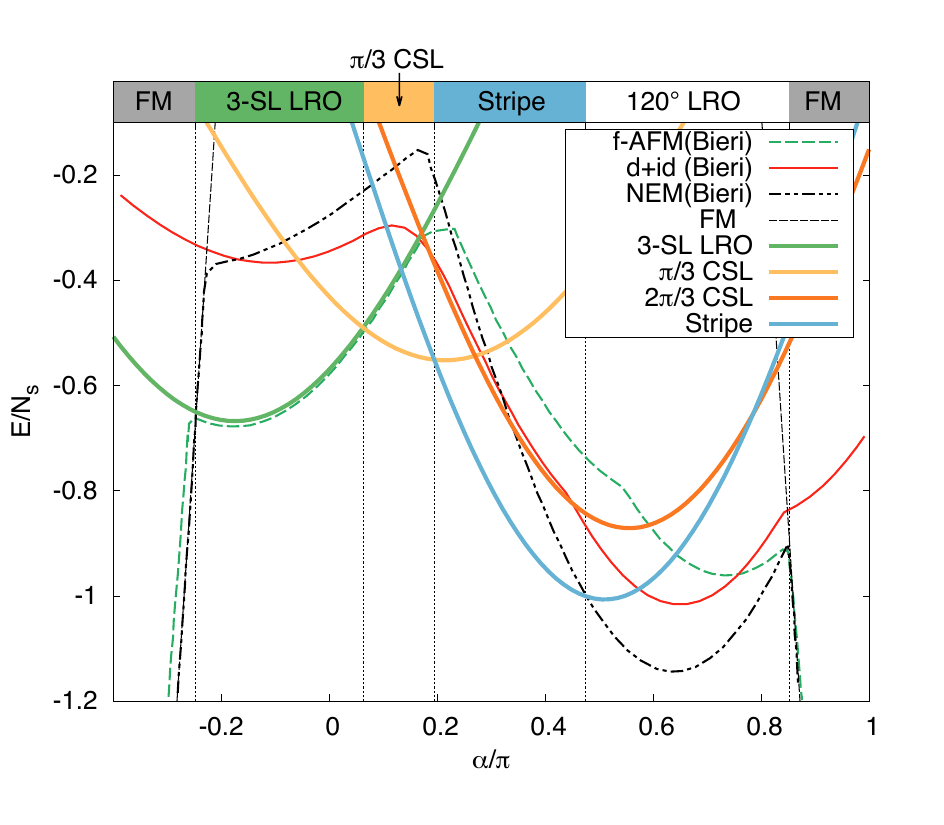}
\caption{Variational energies for the $J$-$K$ model with real $K$ of the most competitive states for the 144-site system compared to the results by Bieri et al.~\cite{Bieri2012} (based on Fig.~3 of their paper).}
\label{fig:Biericomparison}
\end{center}
\end{figure}

As mentioned above, we found three relevant phases in our VMC calculations: a three-sublattice color-ordered state with $0$-flux per plaquette,  the $\pi/3$-flux per plaquette CSL state, and a $\pi$-flux striped phase. We plot the energies of these phases together with the states proposed in Ref.~\cite{Bieri2012} in Fig.~\ref{fig:Biericomparison}. Due to the different focus of their paper, Bieri et al. use a nomenclature more suitable for spin-one systems. However, we will reinterpret their findings in the SU(3) language.

In the coupling regime $-0.248 \pi < \alpha < 0.064\pi$ ($J>0$, \mbox{$-0.99 < K/J < 0.20$}), we find that the energy of the $120^\circ$ AFM order proposed by Bieri et al.~\cite{Bieri2012} and the energy of the 3-SL LRO we propose are almost exactly matched.
In fact, for special values of the parameters used by Bieri et al. their variational states are equivalent to our 3-SL LRO variational states. Namely, for $\sin \eta=  \sqrt{2/3}$ in Eq.~(15) of their paper, the three $\mathbf{d}$ vectors become mutually orthogonal to each other, and their mean-field ansatz can be transformed to ours with an SU(3) rotation. We find that the 3-SL LRO is the more appropriate identification of this phase, as the $120^\circ$ order is just a special case, and in fact not all members of the ground-state manifold have dipole ordering due to the \mbox{SU}(3) symmetry of the model, which can mix the dipolar and quadrupolar moments of the spins 1. 

For $0.064 \pi < \alpha < 0.195 \pi$ ($K>0$, $0.20 < K/J < 0.70$) we find that a $\pi/3$-flux CSL phase is selected. This phase was not considered by Bieri et al.~\cite{Bieri2012}, but was predicted by the mean-field study of Lai~\cite{Lai2013a}.
We find 6 linearly independent states with similar energies by considering states with $\pm\pi/3$ flux per triangle (3 states each) and using twisted boundary conditions (compare Sec.~\ref{sec:methodsVMC}).
This fits the ground-state degeneracy of the CSL breaking time-reversal symmetry spontaneously on a torus with genus $g=1$.
We studied further topological properties by calculating the modular matrices and find a qualitative agreement for the predicted chiral central charge and a quantitative agreement for the anyonic exchange statistics.
Details are given in Appendix~\ref{sec:csltopologocalproperties}.

Beyond the $\pi/3$-flux CSL phase, between $ 0.195\pi < \alpha < 0.475 \pi$, a striped state with stronger bonds along straight chains with $t_{\text{inter-chain}}/t_{\text{intra-chain}} \approx 0.2$ and a uniform $\pi$-flux has the lowest energy, superseeding both the $\text{d}+i\text{d}$ phase proposed by Bieri et al. and the $2\pi/3$-flux CSL proposed by Lai~\cite{Lai2013a}.
 
For $0.475\pi<\alpha<0.85\pi$ ($K>0$, $K/J > 12.7$, and $K/J < -0.51$), the 120-nematic or $\mathcal{J}$-nematic phase proposed by Bieri et al. is clearly dominant; its energy is not matched by any of the states we considered. In their construction the on-site terms before the projection select states with directors in an 'umbrella' configuration (see Eq.~(16) in Ref.~\onlinecite{Bieri2012}), interpolating between a ferroquadrupolar and a $120^\circ$ quadrupolar order. However, due to the $\mbox{SU}(3)$ symmetry of the model there are many other degenerate states, containing not only nematic states, but other states obtained by global $\mbox{SU}(3)$ rotations. Therefore, we find that a $120^\circ$ LRO state is more suitable in this context. 
In the $\mbox{SU}(3)$ language, a $120^\circ$ fully color ordered state is only built out of two colors on every site, while the third color is missing in the system. Interestingly, in this subspace the $J$-$K$ Hamiltonian simplifies since the nearest-neighbor exchange and the ring exchange terms are not independent. Namely, for three sites around a triangle
\begin{equation}
\label{eq:reduction}
\begin{split}
P_{ij} + P_{jk}+P_{ki}= P_{ijk}+P_{ijk}^{-1}+1
\end{split}
\end{equation}
in the two-color subspace, just as in the spin-1/2 case. As a result, at the special point $K=-J/2$, the Hamiltonian for the two-color states is a constant, giving the same energy for any state in this subspace. This macroscopic degeneracy can also be seen in ED results, and the transition between the $120^\circ$ LRO and ferromagnetic $\mbox{SU}(3)$ phases is found quite accurately at \mbox{$\alpha_{\rm c}= \arctan(-1/2)=0.852\pi$} [$K>0$, $(K/J)_{\rm c} > -0.50$] in both ED and VMC. Note that this value of $\alpha$ corresponds to ferromagnetic nearest-neighbor coupling $J<0$, therefore only considering antiferromagnetic $J$, the $120^\circ$ LRO persists for $K\to \infty$.  

Finally, following the argument of Bieri et al., the other end of the ferromagnetic phase can be estimated by comparing the variational energy of the 3-SL LRO state with the energy of a fully polarized ferromagnetic color state, \mbox{$E_{\text{FM}}/N_s= 3 \cos \alpha +4 \sin \alpha$}. This gives a transition at \mbox{$\alpha_{\rm c} = -0.248\pi$} [$K<0$, $(K/J)_{\rm c} = -0.988$].

The phase diagrams for the $J$-$K$ model by ED and VMC, depicted in Fig.~\ref{fig:phasediagramJKmodel}, are in good qualitative agreement.
For small ring exchanges $K$ around $\alpha=0$ we find the 3-SL LRO phase, which is expected in a slightly smaller area for $K>0$ in VMC than in ED.
This is most likely linked to the $\pi/3$-flux CSL at larger ring exchanges $K$, which is particularly well described as a variational state.
The CSL states from VMC and ED show a direct correspondence in terms of symmetries. Details are given in Appendix~\ref{sec:CSL:symmetries}.
In the VMC the next competitive state is the striped state, which is linked to the LN phase in ED, since both break rotational symmetry.
We therefore assume that these phases correspond to the same ground state in the thermodynamic limit, and  stick with the term LN in the following.
The extension of the 120$^\circ$ LRO phase is strongly dependent on the method, whereas the one of the FM is quite similar from ED and VMC.

\section{Hubbard model}
\label{sec_hubbard}
%
We found that the $J$-$K$ model hosts a spontaneous time-reversal symmetry breaking CSL for intermediate values of $K/J\approx 0.35$ for $K>0$ ($\alpha \approx 0.11 \pi$).
Here we turn to the question whether this CSL is also present in the Mott phase of the corresponding Hubbard model \eqref{eq:ham_hubbard}.
In the limit of small values of $t/U$ the Mott phase is well described by the leading nearest-neighbor Heisenberg model stabilizing the 3-SL LRO. With increasing $t/U$, the next subleading interaction, the third-order $K$-term, triggers a phase transition to the CSL. The ED critical value is $(K/J)_{\rm c} \approx 0.31$ ($\alpha_{\rm c} \approx 0.096 \pi$), corresponding to $(t/U)_{\rm c} \approx 0.064$ [$(U/t)_{\rm c} \approx 15.6$] taking the bare third-order series. In the following we show that higher-order contributions do not prevent the occurrence of the CSL phase, and that the critical value remains similar.

\subsection{Effective description}
\label{sec_hubbard_effmodels}
In this section we state and discuss the effective model for the Mott phase of a general SU($N$) Hubbard model at a commensurate 1/$N$ filling for arbitrary uniform fluxes $\Phi$ for the thermodynamic limit.
The model of the 12-site cluster with PBCs, which we use for the direct comparison between the effective description and the Hubbard model in Sec.~\ref{sec:hubbardcomparison}, is given in Appendix~\ref{sec:apptwelvesites}.

In fifth-order perturbation theory the effective model in the thermodynamic limit contains 13 different types of interactions involving permutations on up to five sites.
The effective Hamiltonian reads
\begin{widetext}
\effectivehamiltonian{}
\end{widetext}
where the pictogram underneath every sum illustrates which sites on the full lattice are addressed. This has to be understood as follows. 
The relative angles between the bonds of a graph are fixed and characterize the graph
(e.g.~the graph of $L_{\text{d}}^{3\text{sp}}$ can not be transformed into the one of $L_{\text{s}}^{3\text{sp}}$), however every possibility of rotation has to be included (e.g.~the graph of $L_{\text{d}}^{3\text{sp}}$ can be rotated around the axis defined by $i$ and $j$ by $\pi$). Then, every distinct set of sites contributes to the Hamiltonian.
The prefactors starting in second and third order are
\begin{widetext}
\begin{equation}
\label{eq:Seff_model_constants}
\begin{aligned}
&\epsilon_0 = -6 \frac{t^2}{U^2} - 12 \cos \Phi \, \frac{t^3}{U^3}  + \left( - 26 - 24 \cos 2 \Phi \right) \frac{t^4}{U^4} - \left( 92 \cos \Phi + 60 \cos 3\Phi \right) \frac{t^5}{U^5}\ ,\\
&J = 2 \frac{t^2}{U^2} + 12 \cos \Phi \, \frac{t^3}{U^3}+ \left(12 + 40 \cos 2\Phi \right) \frac{t^4}{U^4} + \left(\frac{820}{9} \cos \Phi + 140 \cos 3\Phi \right) \frac{t^5}{U^5}\ ,\\
&K = - 6 \e{i\Phi} \, \frac{t^3}{U^3} + \left(- 4 - 30 \e{i2\Phi} \right) \frac{t^4}{U^4} - \left( \frac{58}{3} \e{i\Phi} - 16 \e{-i\Phi} + 135 \e{i3\Phi} \right) \frac{t^5}{U^5}\ ,\\
&L^{\text{2sp}}_\text{s} = \frac{10}{3} \frac{t^4}{U^4} + \left( \frac{104}{3} \cos \Phi + 20 \cos 3 \Phi\right) \frac{t^5}{U^5}\ ,\\
&L^{\text{2sp}}_\text{d} = \left( \frac{20}{3} + 8 \cos 2\Phi \right) \frac{t^4}{U^4} - \left( \frac{208}{3} \cos \Phi + 40 \cos 3\Phi \right) \frac{t^5}{U^5}\ ,\\
&L^{\text{3sp}}_\text{s} = -\frac{4}{3} \frac{t^4}{U^4} - \left( \frac{32}{3} \cos \Phi + 30 \cos 3\Phi \right) \frac{t^5}{U^5}\ ,\\
&L^{\text{3sp}}_\text{d} = \left( -\frac{4}{3} - 10 \e{i2\Phi} \right) \frac{t^4}{U^4} - \left( \frac{32}{3} \cos \Phi + \frac{224}{9} \e{i\Phi} + 60 \e{i3\Phi} \right) \frac{t^5}{U^5}\ , \\
&L^{\text{4sp}}_\text{r} = 20 \e{i2 \Phi} \frac{t^4}{U^4} + \left( \frac{232}{9} \e{i\Phi} + 140 \e{i3\Phi} \right) \frac{t^5}{U^5}\ ,\\
&L^{\text{3sp}}_\text{cr} = - \left( \frac{112}{9} \e{i\Phi} + 15 \e{i3\Phi} \right) \frac{t^5}{U^5}\ , \qquad
L^{\text{4sp}}_\text{cr} = \frac{58}{9} \e{i\Phi} \, \frac{t^5}{U^5}\ , \qquad
L^{\text{4sp}}_{\text{pl}} = \frac{116}{9} \cos \Phi \, \frac{t^5}{U^5}\ ,\\
& L^{\text{4sp}}_\text{Ka} = - \frac{232}{9} \cos \Phi \, \frac{t^5}{U^5}\ , \qquad
L^{\text{4sp}}_\text{Kb} = -\frac{116}{9} \cos \Phi \, \frac{t^5}{U^5}\ , \qquad \text{and} \qquad L_\text{r}^{\text{5sp}} = - 70 \e{i3\Phi} \, \frac{t^5}{U^5}\ .
\end{aligned}
\end{equation}
\end{widetext}

%
%
\begin{figure}[b]
	\centering
		\includegraphics[width=0.63\columnwidth, angle = 90, trim={3.25cm 10cm 7.8cm 2.25cm},clip]{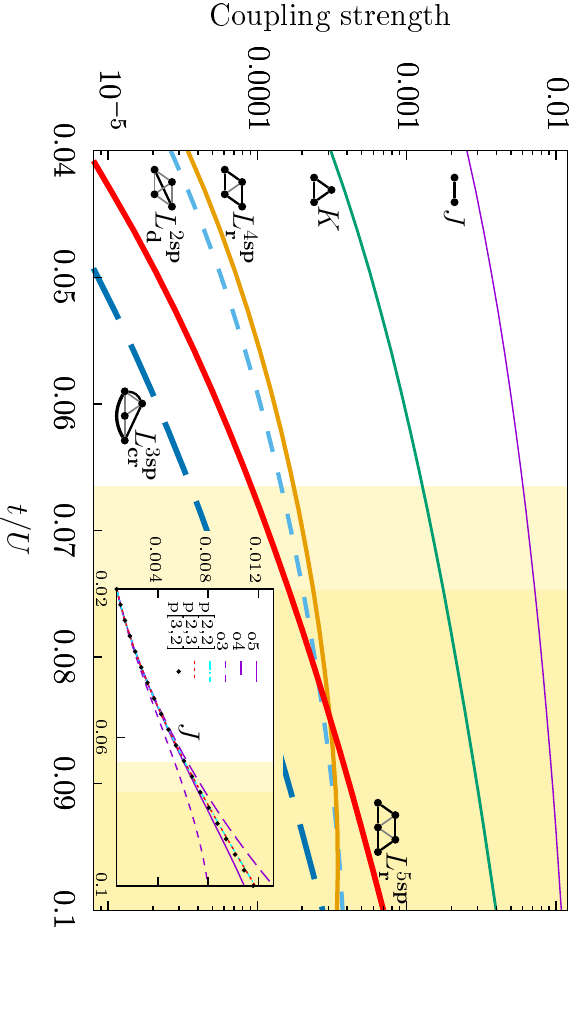}
\caption{Effective couplings in units of $U$ of the Hubbard model as a function of $t/U$ for $\Phi=\pi$ using bare fifth-order series. Plotted are the largest contributions in every order with a pictogram sketching the associated permutations. The dark yellow background indicates the area where the CSL is observed within ED and VMC, whereas the light yellow corresponds to its stability according to VMC only (compare Sec.~\ref{sec:hubbardCSL}).
The inset shows the nearest-neighbor exchange in different orders and Pad\'e extrapolations.}
\label{Fig:couplings_pi}
\end{figure}

%
\begin{figure}[b]
\centering
\includegraphics[width=1\columnwidth]{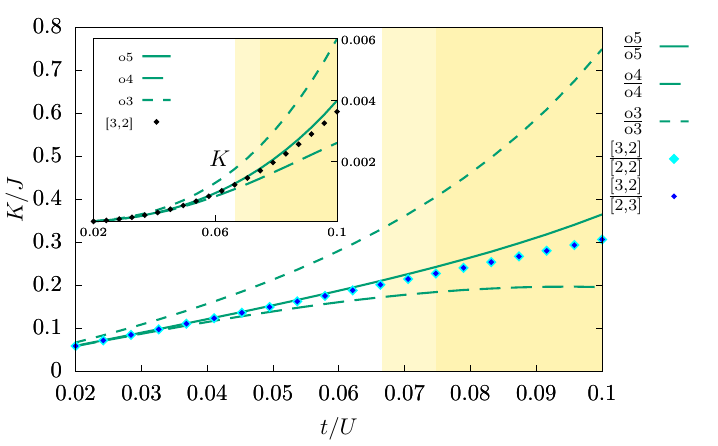}
\caption{Ratio of effective coupling constants $K/J$ depending on $t/U$ for $\Phi=\pi$ using bare series up to order 5 (green) as well as Pad\'e extrapolations (cyan, blue). The inset shows a similar plot for the three-site ring exchange $K$ in units of $U$ by itself. The background colors are defined as in Fig.~\ref{Fig:couplings_pi}.}
\label{Fig:S_coupling_convergence_pi}
\end{figure}
%

\begin{figure*}[t]
\centering
\includegraphics[width=\textwidth]{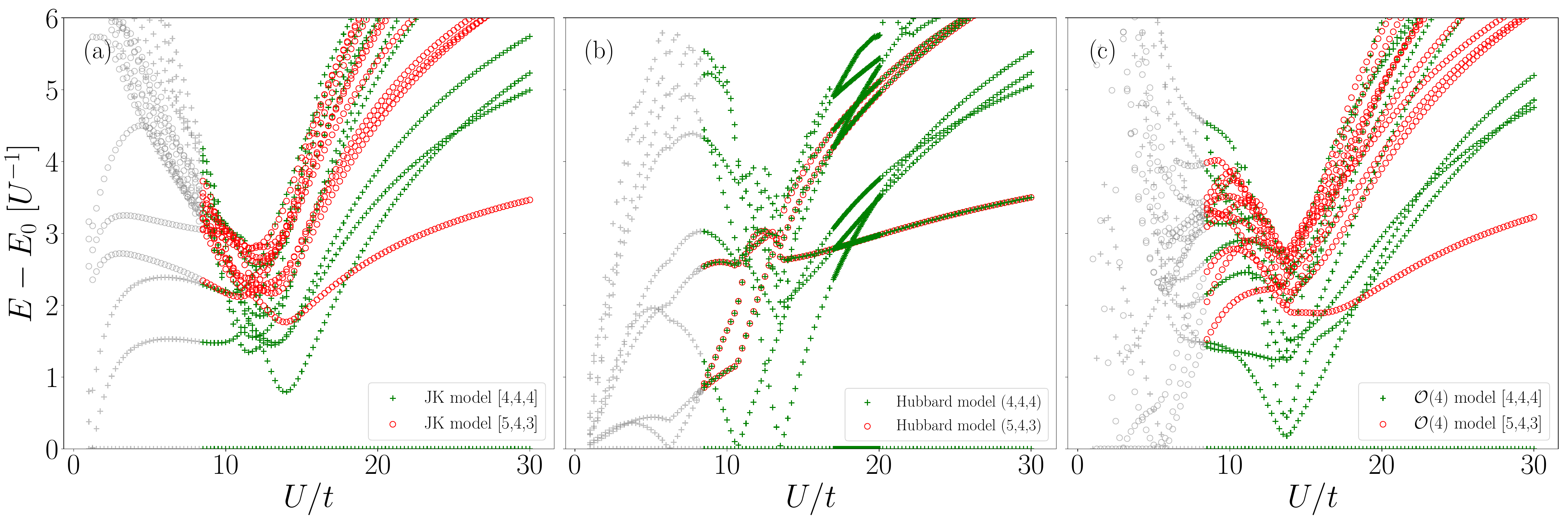}
\caption{Comparison of the excitation spectra between the $J$-$K$ model (left), the Hubbard model (middle) and the $\mathcal{O}(4)$ (order 4) effective model (right) on the 12-site cluster from ED. Since the couplings in the effective descriptions are polynomials in $t/U$, the energies from the spin models are multiplied by $U$ to be comparable to the energies of the Hubbard model. For $U\approx 30t$ the spectra agree very well, but differences become noticeable as $U$ decreases. The grey regions of the spectra are not in the Mott-insulating phase of the Hubbard model and the effective models are not valid.}
	\label{Fig:S_Spin_vs_hubbard}
\end{figure*}

We note that the complex phases of ring exchanges in subleading orders of the effective interaction are not identical to the flux through the apparent plaquettes in the Hubbard model.
For example, the complex phase of a cyclic permutation on a triangle only equals the flux through a triangle in order 3.
The representation of the effective model given above is not unique for real exchange constants if $N<5$.
This is due to the fact that consecutive permutations acting on more sites than spin colors in the model can be rewritten in terms of sums of various other permutation operators.
The best known example occurs for the three-site permutation acting on SU(2) spins, which can be expressed by two-site permutations plus a constant [compare Eq.~\eqref{eq:reduction}].
The situation becomes much richer if one considers permutations between more spins. 
For instance, in the case of SU(3) spins, the four-site permutation can be rewritten in terms of two-spin, three-spin and various four-spin interactions
\begin{equation}
\begin{aligned}
\label{Eq_4sites_SU3}
P_{1234} &+ P_{4321} = 1 - \sum_{(i,j)} P_{ij} + \sum_{(i,j,k)} P_{ijk}+ P_{12}P_{34}\\
&+ P_{14}P_{23} + P_{13}P_{24} - \left(P_{1243} + P_{1324} + \text{h.c.} \right)\ ,
\end{aligned}
\end{equation}
where the sums include all possible permutations of different sites.
If we use this relation to reexpress the four-site ring exchange on a plaquette in the effective Hamiltonian all exchange constants of the exchanges appearing on the right hand side of Eq.~\eqref{Eq_4sites_SU3} get rescaled. Additionally, new interactions occur which in perturbation theory arise only in higher orders. In this sense the replacement of operators is not helpful and the formulation in Eq.~\eqref{eq:H_eff_triangular} is the more natural one in terms of perturbation theory. If and how a systematic reduction of higher-order interactions to only already included exchanges is possible for orders larger than 5 remains an open question.

In the following, we discuss the most interesting case \mbox{$\Phi=\pi$}, where the convergence of the series works particularly well. Additional results for other fluxes are presented in Appendix~\ref{sec:CSLnonpiflux}.
The dependence of the largest coupling constants of every order as bare series in $t/U$ is illustrated in Fig.~\ref{Fig:couplings_pi}. The most important subleading corrections to the nearest-neighbor Heisenberg term come from ring exchanges around triangles, squares, and 5-site trapezoids.
The shaded background keeps track of where the CSL is observed in only VMC (light yellow) and in ED and VMC (darker yellow) (compare Sec.~\ref{sec:hubbardCSL}).
The Pad\'e-extrapolants for the dominant nearest-neighbor coupling $J$ and the subleading three-site ring exchange $K$ are shown in the insets of Fig.~\ref{Fig:couplings_pi} and Fig.~\ref{Fig:S_coupling_convergence_pi}, respectively.
The relevant ratio $K/J$ is plotted in Fig.~\ref{Fig:S_coupling_convergence_pi}, where we take the ratio of the extrapolations of $J$ and $K$.
In respect to unphysical divergences, we found a composition of [3,2]-Pad\'e extrapolations for $\epsilon_0$, $J$, and $K$, [2,1]-Pad\'e extrapolations for
$L_{\text{s}}^{\text{2sp}}$, $L_{\text{d}}^{\text{2sp}}$, $L_{\text{s}}^{\text{3sp}}$, $L_{\text{d}}^{\text{3sp}}$, and $L_{\text{r}}^{\text{4sp}}$, and all other couplings as bare series to work best for flux $\Phi=\pi$.
%
%
%

\begin{figure}[b]
	\centering
		\includegraphics[width=0.9\columnwidth]{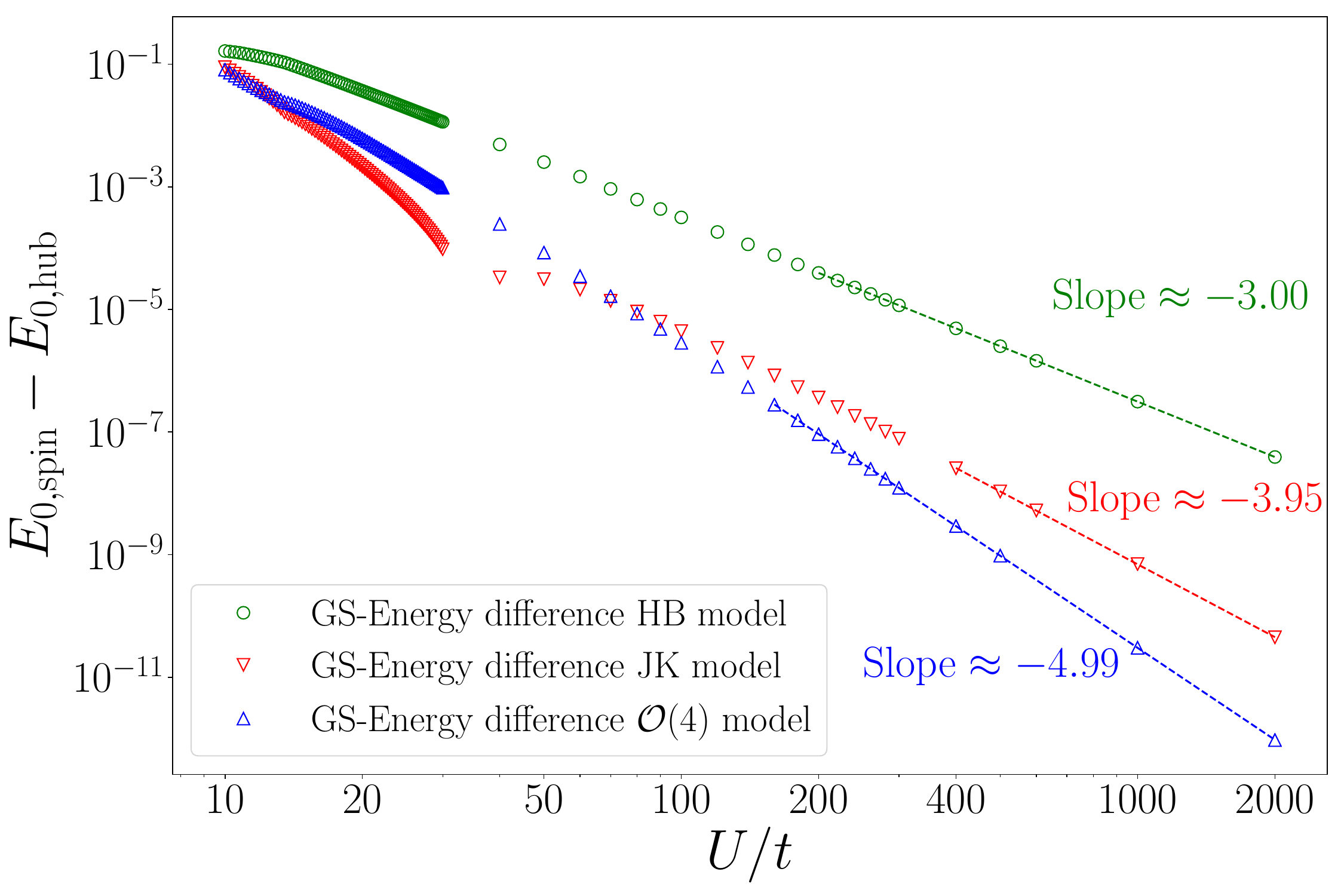}
	\caption{Scaling of the differences in the ground-state energies between various spin models  and the Hubbard model on the 12-site cluster from ED. The effective second order Heisenberg model is abbreviated by HB model. A line is fitted for each spin model to estimate the scaling behavior. The slopes are in good agreement with the expectations.}
	\label{Fig:S_error_scaling}
\end{figure}

\subsection{Comparison}
\label{sec:hubbardcomparison}
In order to confirm the validity of the effective description, we compare the spectrum of the SU(3) Hubbard model \eqref{eq:ham_hubbard} with the spectra of the $J$-$K$ model \eqref{eq:ham_JK} and the $\mathcal{O}(4)$ (order 4) effective spin model \eqref{Eq_H_eff12sites_triangular} on the 12-site cluster (compare Appendix~\ref{sec:apptwelvesites}) from ED.
The results are shown in Fig.~\ref{Fig:S_Spin_vs_hubbard}.
While for the spin models the numbers in the square brackets label a certain irrep of the SU(3) group, the three numbers in the round brackets for the Hubbard model label the number of particles of a certain color, and are thus U($1$) quantum numbers. For $U\approx 30t$ the spectra of both effective models are in reasonable agreement with the spectrum of the  Hubbard model. The first excitation in the spin models is in the adjoint representation $[N_s/3+1,N_s/3,N_s/3-1]$ followed by three singlet levels. For decreasing couplings $U$, the higher lying singlets start to cross each other in the spin models, as the corresponding excited states do in the Hubbard model. In the effective models at $U\approx15-20t$, the first excited singlet crosses the low-energy state of the adjoint representation, which for even smaller values of $U$ is also crossed by two more singlets. One can observe a similar behavior in the spectrum of the Hubbard model, even though the order in which the crossings occur is not exactly the same. In general, the level crossings of the $\mathcal{O}(4)$ model seem to match those of the Hubbard model slightly better than those of the $J$-$K$ model, as expected.

To further check the effective description, we compare the differences between the ground-state energies of the spin models with those of the Hubbard model as can be seen in Fig.~\ref{Fig:S_error_scaling}. The errors of the ground-state energies decay with one order higher than the corresponding order of the effective description, as expected for a valid perturbative approach. Overall, there is a qualitative agreement between the effective description and the Hubbard model in the strong-coupling Mott regime.

\begin{figure}[t]
	\centering
		\includegraphics[width=1\columnwidth]{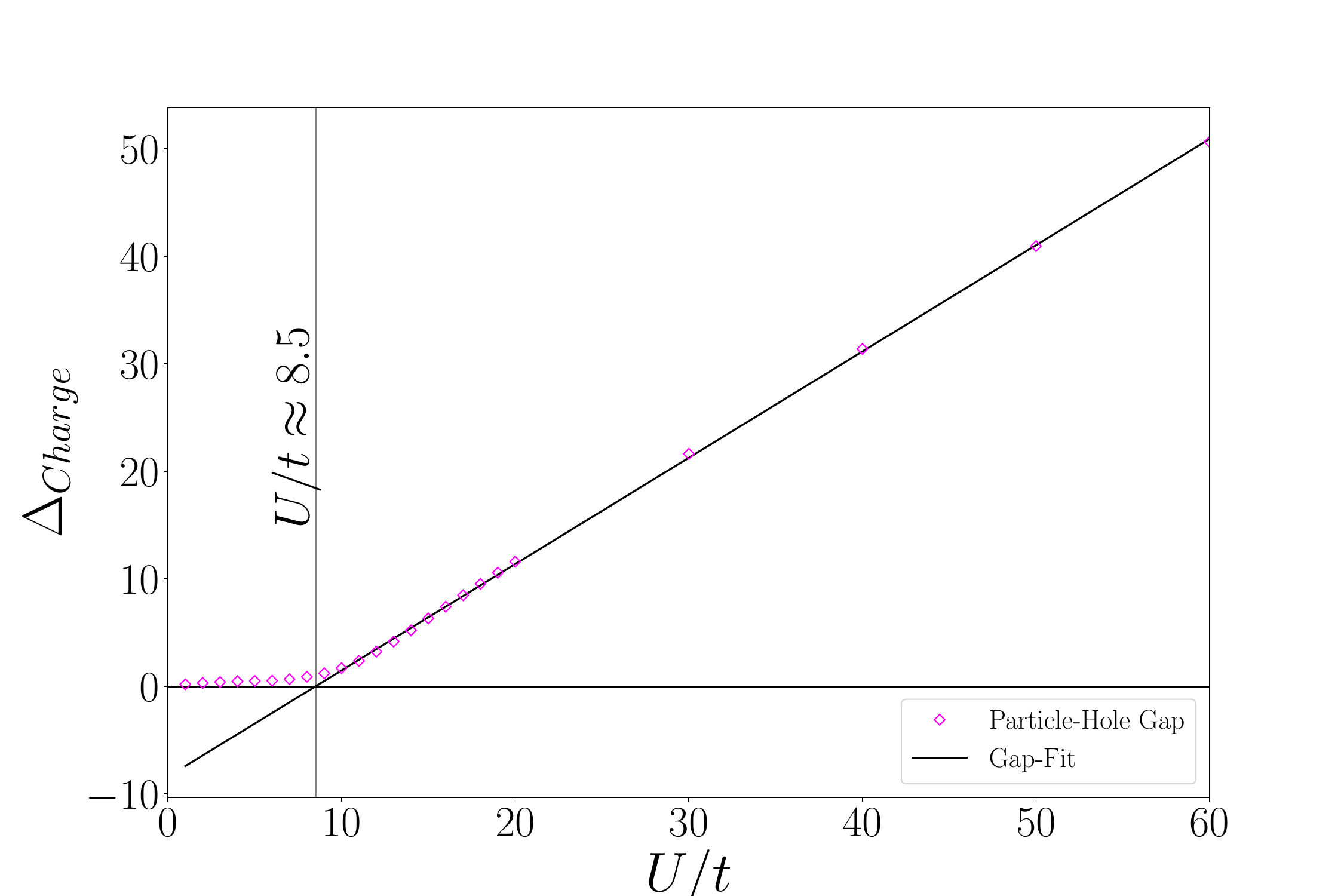}
	\caption{Particle-hole gap for the SU(3) Hubbard model with $\Phi=\pi$ from ED on the 12-site cluster indicating the estimation of the metal-insulator transition.}
	\label{Fig:mott_transition_estimate}
\end{figure}

An important aspect of the problem is that the effective description breaks down at the metal-insulator transition.
Therefore, the physics found in the effective spin model is only valid in the Mott phase and it is important to estimate the metal-insulator transition point.
To this end, we compute the particle-hole charge gap for the $\pi$-flux SU(3) Hubbard model on the 12-site cluster. The results are shown in Fig.~\ref{Fig:mott_transition_estimate}. They indicate that the metal-insulator transition is located at $(U/t)_{\rm c}^{\rm mi}\approx 8.5$.
This value is further discussed at the end of the next Subsection, which mainly focuses on the presence of the spontaneous time-reversal symmetry broken CSL in the fifth-order effective model within the Mott phase.

\subsection{Chiral spin liquid}
\label{sec:hubbardCSL}
%
Next, we analyze the effective spin model in Eq.~$\eqref{eq:H_eff_triangular}$ for the specific case $N=3$ with $\pi$-flux on the triangular lattice.
For small values of $t/U$ the couplings up to third order are dominant, hence the $J$-$K$ model is well converged.
The phase transition between the 3-SL LRO and the CSL occurs at \mbox{$(K/J)_{\rm c} \approx 0.31$}, which translates to $(t/U)_{\rm c} \approx 0.09$ [$(U/t)_{\rm c} \approx 11$] in bare fifth-order.
Interestingly, this critical value changes only slightly to $(t/U)_{\rm c} \approx 0.1$ [$(U/t)_{\rm c} \approx 10$] when applying Pad\'e extrapolations to the couplings $J$ and $K$. For both couplings several Pad\'e extrapolants give essentially the same result in this $t/U$-regime so that the extrapolations work well for the most important interactions of the effective model (see also the inset in Fig.~\ref{Fig:couplings_pi}). The small difference between the critical ratios from bare series and extrapolation within the $J$-$K$ model indicates that even the bare series is almost converged up to these $t/U$ values for the couplings $J$ and $K$.

%
%
\begin{figure}[t]
\centering
\includegraphics[width=1\columnwidth]{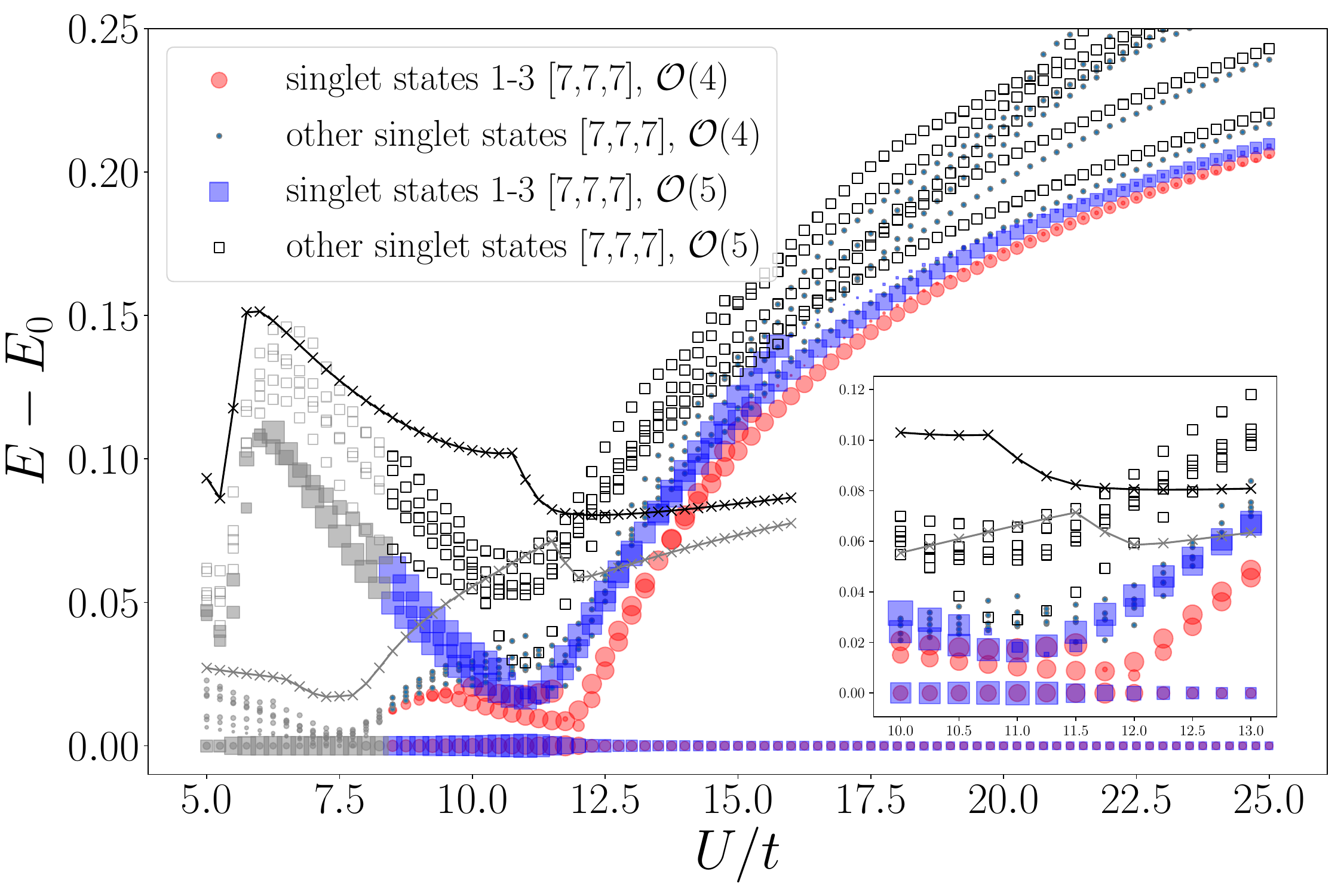}
\caption{Spectrum of the fourth- and fifth-order effective model for $\Phi = \pi$ from ED on the 21-site cluster.
The marker size corresponds to the overall chirality signal and is plotted for the lowest three states. Three low-lying singlets with strong chirality signal indicate the presence of a CSL phase around $U/t \approx 12$.
The grey regions of the spectra are not in the Mott-insulating phase of the Hubbard model and the effective models are not valid. The grey (black) solid line indicates the lowest energy eigenstate in the adjoint irrep
[8,7,6] of the fourth- (fifth-)order effective model. The inset is a zoom into the CSL region.}
\label{Fig:spectrum_ED_4thand5thorder}
\end{figure}

\begin{figure}[t]
\centering
\includegraphics[width=1\columnwidth]{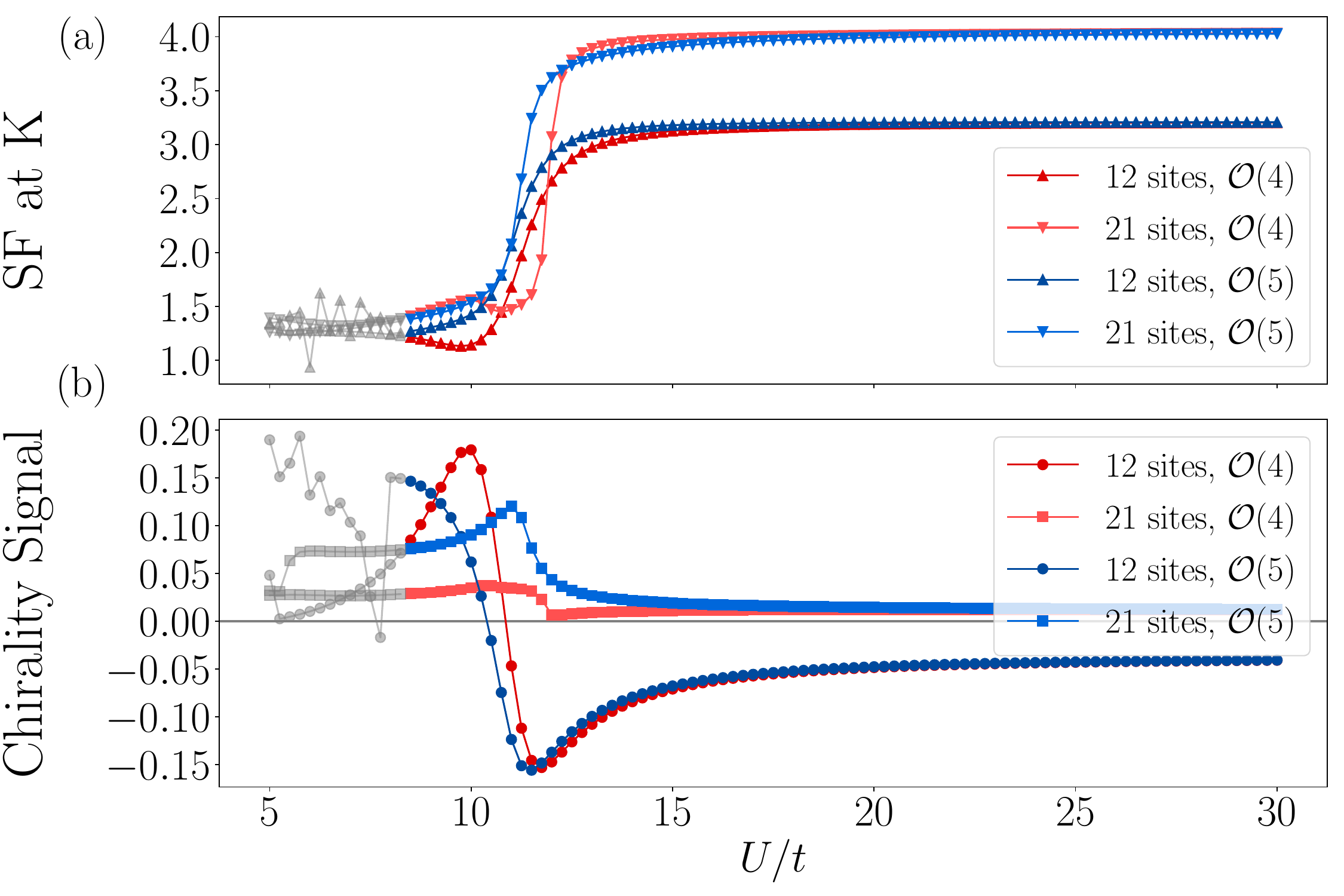}
\caption{Structure factor at the K point (top) and chirality signal per lattice site (bottom) for the fourth- and fifth-order effective model for $\Phi = \pi$ from ED on the 21-site cluster.
Coming from the 3-SL LRO phase at large ratios $U/t$, the structure factors at the ordering momentum K decrease as the chirality signals increase around $U/t \approx 12$. The grey regions of the structure factors and the chirality signals
are not in the Mott-insulating phase of the Hubbard model and the effective models are not valid.}
\label{Fig:observables_ED_4thand5thorder}
\end{figure}

Since the impact of all the other smaller terms is not obvious a priori, we explicitly study them with the set of Pad\'e extrapolants for the coupling constants discussed in Sec.~\ref{sec_hubbard_effmodels} using ED and VMC.
The low-energy spectra of the fourth- and fifth-order effective model for flux $\Phi=\pi$ on the 21-site cluster from ED are illustrated in Fig.~\ref{Fig:spectrum_ED_4thand5thorder}, where again the point size for the lowest three singlets illustrates the chirality signal.
For large values of $U/t$ the tower of states characteristic of the 3-SL LRO phase appears in the spectrum. This is in agreement with the large structure factor and its extensive scaling at the K point, given in the upper panel of Fig.~\ref{Fig:observables_ED_4thand5thorder}.
In the fifth-order model the tower disappears at $(U/t)^{\text{ED},\mathcal{O}(5)}_c \lesssim 13$ [$(t/U)^{\text{ED},\mathcal{O}(5)}_c \gtrsim  0.075$], where three low-lying singlet levels occur with the same degeneracies 1-4-1 as in the CSL of the $J$-$K$ model (compare Fig.~\ref{Fig:S_Spectra_chiral_states}).
The direct correspondence between the states in the effective models is also clear from the symmetries discussed in Appendix~\ref{sec:CSL:symmetries}.
Therefore, the CSL is most plausibly present here as well.
In the same parameter range $U/t \approx 13$ the chirality signal of the ground state increases, as can be seen in the bottom panel of Fig.~\ref{Fig:observables_ED_4thand5thorder}, whereas the structure factor at the K point decreases.
This behavior agrees perfectly with the phase transition from the 3-SL LRO phase to the CSL.
For values $(U/t)^{\text{ED},\mathcal{O}(5)}_c \lesssim 11$ [or $(t/U)^{\text{ED},\mathcal{O}(5)}_c \gtrsim  0.09$] another state drops down, and the signature of six low-lying states is lost. However, the chirality signal in the lowest states remains large.

Within the area of the potential CSL the manifold of six low-lying states is not very well separated from the excited states, but the indications are stronger on the 21-site cluster than on the 12-site one (not shown). Since ED including all exchanges is a lot more costly than for the $J$-$K$ model, we did not study the 27-site cluster.
However, with VMC on the 21-site cluster we find the same phase transition from the 3-SL LRO to the $\pi/3$-flux CSL at $(U/t)^{\text{VMC},\mathcal{O}(5)}_c\approx 14.9$ [$(t/U)^{\text{VMC},\mathcal{O}(5)}_c\approx 0.067$].
The CSL states behave very similar to the lowest eigenstates from ED regarding energy spectra and symmetries.
All these findings strongly point to the realization of a $\pi/3$-flux CSL phase with spontaneous breaking of time-reversal symmetry.

The energy spectra of the fourth- and fifth-order model from ED in Fig.~\ref{Fig:spectrum_ED_4thand5thorder} behave fairly similar for $U/t \gtrsim 10$.
For large ratios $U/t$ the eigenenergies approach each other, as expected.
With increasing perturbation (decreasing $U/t$) the differences increase.
Nevertheless, the same manifolds of six low-lying states emerge.
Also on the level of observables the signature of the CSL is present in both models as shown in Fig.~\ref{Fig:observables_ED_4thand5thorder}.
So, even though the effective model is not quantitatively converged in the relevant area $11 \lesssim U/t \lesssim 13$, the signature of the CSL occurs in the third-, fourth-, and fifth-order model, which implies, that it is a definite feature of the effective description of the Hubbard model in this $U/t$ regime.
For coupling ratios $U/t\lesssim 10$ the behavior of the eigenstates is rather different in the fourth- and fifth-order model, so no statements about this parameter range can be made, apart from the fact that the Mott phase breaks down eventually.

The question is whether the potential CSL phase is within the Mott phase of the SU(3) Hubbard model with flux $\Phi=\pi$. In the ED calculations on 12 sites, the charge gap (given in Fig.~\ref{Fig:mott_transition_estimate}) shows a linear behavior above $U\approx 10|t|$, which together with the effective model (compare Appendix~\ref{sec:apptwelvesites}) provides the estimate $(U/t)^{\text{mi}}_{\rm c} \approx 8.5$.
Can we interpret this result in first-order perturbation theory around the limit of strong coupling at filling 1/3?
Assuming linear behavior for the charge gap the transition point yields \mbox{$\Delta_{\text{charge}} \approx U- 8.5 t$}, and the first order term must arise from the kinetics of charge excitations.
These are given by doublon-hole pairs. In the case of the SU(2) model, both the hole and the doubly occupied site are featureless objects, in fact SU(2) singlets, and they behave similarly. In the case of the SU(3) Hubbard model, the doubly occupied site forms the three-dimensional anti-symmetrical irreducible representation, and the motion of the doubly occupied site is more complicated than the motion of the hole. The energy $-8.5 t$ originates from the hoppings of the hole and the doubly occupied sites, the contribution from the hole is \mbox{$E_0(-1)- E_0(0) \approx -0.02 U - 3.30t$,} and the contribution of the doubly occupied site is $E_0(+1)- E_0(0) \approx 1.01 U - 5.10t$ [c.f. Eq.~(\ref{eq:Deltacharge})].
We can see, that the kinetic energy of the doubly occupied site is larger than that of the hole.  

The metal-insulator transition from ED on the 12-site cluster at $(U/t)_{\rm c}^{\text{mi}} \approx 8.5$ [$(t/U)_{\rm c}^{\text{mi}} \approx 0.12$] occurs for weaker coupling strengths $U$ than the phase transition towards the CSL.
For the $J$-$K$ model the estimated transition point between 3-SL LRO and CSL is located at slightly larger values $(U/t)_{\rm c} \approx 10$, and therefore lies within the crudely estimated extension of the Mott phase.
However, the fifth-order effective model includes a larger variety of quantum fluctuations and is therefore more reliable with the CSL below $(U/t) \lesssim 13$.
We thus expect that the Mott phase of the SU(3) $\pi$-flux Hubbard model on the triangular lattice realizes, besides the 3-SL LRO phase, a spontaneous time-reversal symmetry broken CSL phase
before the Mott insulating phase breaks down.
The full phase diagram for the Hubbard model is illustrated in Fig.~\ref{fig:phasediagramHubbardmodel}.

%
%
\section{Conclusion and outlook}
\label{sec:outlook}
%
%
  To summarize, we have provided very strong numerical evidence that the stabilisation of a chiral phase can be achieved with cold atoms in a simple and realistic setting: three flavours (e.g.~$^{87}$Sr or $^{173}$Yb) and a triangular optical lattice.\\ 
  To achieve this, we have first studied the SU(3) J-K model on the triangular lattice and shown that it has a CSL phase that spontaneously breaks time-reversal symmetry for a positive but moderate value of three-site exchange term $K$. In the search for realistic models with a CSL phase that could be implemented experimentally, this is an important step forward because the two interactions that appear in this model, the nearest-neighbor exchange and the three-site permutation around triangles, are the most relevant ones in the Mott phases of the model when coming from strong coupling. So in principle it is accessible to cold atom experiments on SU(3) fermions by just monitoring the ratio of the repulsion between fermions $U$ to their hopping amplitude.


To make this proposal more concrete, we have discussed in great details the connection between the Hubbard model and the $J$-$K$ model with positive $K$. For the physically most relevant sign of the hopping, $-t$ with $t>0$ to get a band with a quadratic spectrum around its bottom at zero momentum, a positive three-site term $K$ can be reached in two situations: either in the Mott phase with one particle per site if one introduces a $\pi$ flux per plaquette to effectively change the sign of the hopping,  or in the Mott phase with two particles per site without any flux since it is equivalent to the Mott phase with one hole per site starting from the full system, and the sign of the hopping changes under a particle-hole transformation.

It is also important to check if, upon reducing the repulsion $U$, the three-site term gets large enough to induce the transition from the three-sublattice color order into the CSL before the Mott transition into a metallic phase takes place. To get an estimate of this Mott transition, we have investigated the Hubbard model directly on small clusters, and indeed the three-site interaction reaches values large enough to be in the CSL phase before the system turns metallic.

Finally, to lend further support to this proposal, we have pushed the strong-coupling expansion for the Mott phase to higher order to check the effect of residual interactions, and the conclusion is that a CSL phase appears to survive the inclusion of these terms.

Altogether, the proposal that there is a CSL phase in the SU(3) Hubbard model on the triangular lattice between the three-sublattice color ordered phase and the metallic phase is we believe fairly solid, and we hope that the present paper will encourage the experimental investigation of this model with cold fermionic atoms.

In that respect, we would like to briefly discuss the temperature effects. Since the chiral phase reported here spontaneously breaks time-reversal symmetry, it is expected to give rise to an Ising transition at finite temperature. It would be nice to have an estimate of this Ising temperature to see if it is accessible for the lowest entropies per particle that can be reached with state-of-the-art cooling protocoles. Unfortunately our methods are limited to zero temperature, and this has to be left for future investigation.

Finally, as a byproduct, we have also come up with a new version of the full phase diagram of the $J$-$K$ model with real $K$, including all signs of $J$ and $K$, and we have shown that, just above the CSL, it contains a large lattice nematic phase that had been overlooked so far. Whether such a phase is also present in the Hubbard model before the Mott transition remains to be seen. 

\section*{Acknowledgement}

We acknowledge financial support by the German Science Foundation (DFG) through the Engineering of Advanced Materials Cluster of Excellence (EAM) at the Friedrich-Alexander University Erlangen-N\"urnberg (FAU), by the Swiss National Science Foundation, by the Hungarian NKFIH Grant No. K124176, and by the BME-Nanonotechnology and Materials Science FIKP grant of EMMI. CJG and AML thank the Austrian Science Fund FWF for support within the project DFG-FOR1807 (I-2868).
The  computational results presented have been achieved using the HPC infrastructure LEO of the University of Innsbruck, the MACH2 Interuniversity Shared Memory Supercomputer,  the facilities of the Scientific IT and Application Support Center of EPFL, and the Erlangen Regional Computing Center (RRZE).


\appendix

\section{Details on ED using irreps of SU($N$)}
\label{sec:detailsonED}
The method described in Ref.~\onlinecite{Nataf2014} allows to directly construct a basis of a given irrep of the SU(3) group and to compute the matrix elements of the Hamiltonian efficiently.
In general, any SU(3) invariant Hamiltonian which can be expressed solely in terms of transposition operators $P_{ij}$, see Eq.~\eqref{eq:permutationoperator}, can be effectively dealt with using such a Young tableau basis.
A general Young tableau for SU($N$) has a maximum of $N$ rows.
For $N_s$ sites, each hosting a particle in the fundamental representation, the possible Young tableaus contain $N_s$ boxes in total.
That means, it can be labeled by $N$ integers [$n_1,n_2,\hdots,n_N$], where $n_i$ counts the number of boxes in the $i$-th row. According to the rules of Young tableau diagrams, the number of boxes in each row must not increase going from top to bottom.
As an example, Fig. \ref{fig:youngtableaux} shows all possible SU(3) Young tableau diagrams for $N_s=3$ sites. Each of these diagrams corresponds to an irrep of SU(3).
Thus, for any given $N_s$ we can draw the Young-diagram representing the irrep of SU(3) we want to consider, and construct the basis for this irrep as well as the matrix elements of the Hamiltonian using the rules presented in Ref.~\onlinecite{Nataf2014}.

\begin{figure}[t]
\begin{center}
 \yng(3) [$3,0,0$] \hfill
 \yng(2,1) [$2,1,0$] \hfill
 \yng(1,1,1) [$1,1,1$]
\caption{All possible SU(3) Young tableaus for $N_s=3$ and the corresponding labels.}
\label{fig:youngtableaux}
\end{center}
\end{figure}

\begin{figure}[h]
\begin{center}
\includegraphics[width=\linewidth]{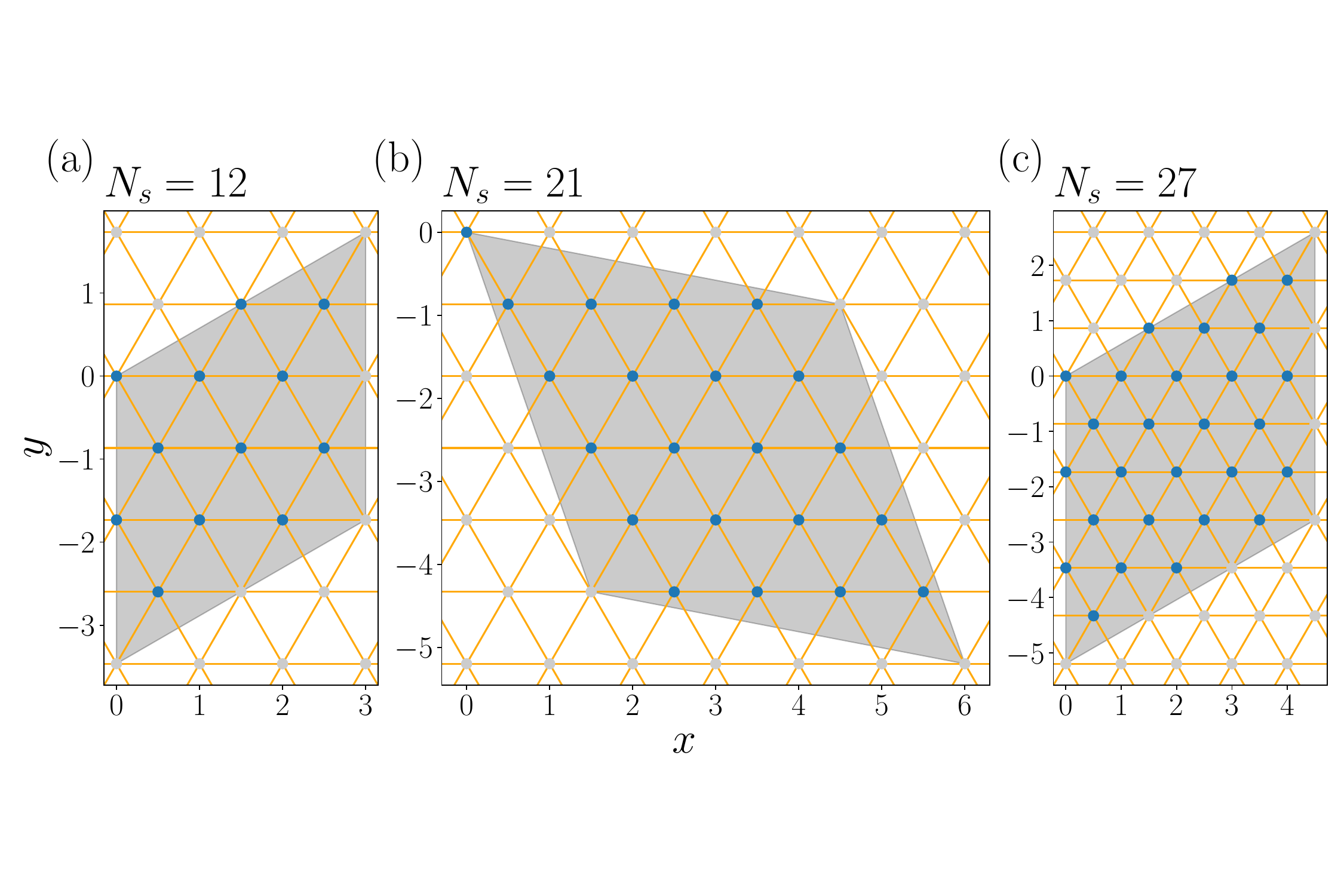}
\caption{Triangular lattice clusters with 12-, 21-, and 27-sites used to derive the ED results presented in this work.}
\label{fig:edclusters}
\end{center}
\end{figure}

Fig.~\ref{fig:edclusters} shows the different clusters used to derive the ED results.

\section{Details on the LCE}
\label{sec:detailsonLCE}
Let us consider a generic problem on an arbitrary lattice
and an extensive quantity $P$. Then, the ratio $P/n$ with the number of lattice sites $n$ is given by a weighted sum over all topological different clusters $c$ 
\begin{equation}
\label{Eq_LCE}
\frac{P}{n} = \sum_c L(c) W_P(c)\ .
\end{equation}
The multiplicity $L(c)$ is the number of ways in which a cluster can be embedded on the
full lattice of interest and determines the impact of each cluster.
The weight of a cluster $W_P(c)$ to the property $P$ is given by the inclusion-exclusion principle. This means that only reduced contributions appearing on the considered cluster but not on its subclusters are taken into account
\begin{equation}
\label{inclusionexclusion}
W_P(c)=P(c) - \sum_{s\subset c} W_P(s)\ .
\end{equation}
One calculates the physical property $P(c)$ on cluster $c$ and subtracts the weights of all subclusters $s$.
Evidently, all processes on a disconnected cluster are already included in the sum of its pieces and the weight vanishes. Therefore, only linked clusters contribute and the weight of a cluster contains only properties that arise from all bonds of the cluster.
This fact can be used to calculate the weights directly with a white-graph expansion~\cite{Coester2015}, which makes the subtraction in Eq.~\eqref{inclusionexclusion} unnecessary.
To this end, every bond on a linked-cluster is labeled with a different exchange constant during the calculation. For instance, on the triangle we take three exchange constants $h_1$, $h_2$, and $h_3$ connecting different sites. The perturbation is written as
\begin{align*}
\mathcal{V}_{\text{triangle}} = \sum_{\alpha =1}^3 \left( h_1  c_{1\alpha}^\dagger c_{2\alpha}^{\phantom{\dagger}} + h_2 c_{2\alpha}^\dagger c_{3\alpha}^{\phantom{\dagger}} + h_3 c_{3\alpha}^\dagger c_{1\alpha}^{\phantom{\dagger}} \right) + \text{h.c.}\ .
\end{align*}
The subtraction of the effective Hamiltonian derived in this way is then achieved by taking only terms that include every exchange constant at least once and hence emerge from perturbations that link the whole cluster.
This procedure can be extended to include complex phase factors by splitting up a hopping process on a link into the hopping from left to right and from right to left. For the triangle one can choose
\begin{align*}
\mathcal{V}_{\text{triangle}} &= \sum_{\alpha = 1}^3 \left( h_{1A}  c_{1\alpha}^\dagger c_{2\alpha}^{\phantom{\dagger}} + h_{1B} c_{1\alpha}^\dagger c_{2\alpha}^{\phantom{\dagger}} \right)\\
&+ \sum_{\alpha =1}^3 \left( h_2 c_{2\alpha}^\dagger c_{3\alpha}^{\phantom{\dagger}} + h_3 c_{3\alpha}^\dagger c_{1\alpha}^{\phantom{\dagger}} \right) + \text{h.c.}
\end{align*}
with $h_{1A} = h_{1} \e{\ci\Phi} = h_{1B}^*$.


\section{Eigenvalues of the Quadratic Casimir Operator for SU(2) and SU(3)}
\label{app:quadraticcasimir}
In Fig.~\ref{Fig:S_TOS_SU2} it seems that every eigenvalue of the SU(2) quadratic Casimir operator is shifted to larger values by some offset, but reappears in the SU(3) observable. Since this is not obvious, we determine the origin of this effect in the following.
To this end, we first show that the lowest eigenvalue of the SU(2)-like irreps inside SU(3) (the irreps with only two colors) depends only on the number of lattice sites $N_s$.
Second, we prove that the eigenvalues have the same slope as their SU(2) counterparts.
We start from the SU(3) case, and consider
an irrep $[n_a,n_b,n_c]$, which can also be labeled by only two numbers
\begin{align}
\label{eq: p and q}
 p &= n_a-n_b\ , \\
 q &= n_b-n_c\ .
\end{align}
In the language of Young tableaus $p$ and $q$ give the difference in the numbers of columns between the first and second row, and the second and third row, respectively.
The eigenvalues of the quadratic Casimir operator in the SU(3) case are then given by
\begin{align}
\label{eq: Quadratic casimir SU3}
 C_2^{\text{SU}(3)}(p,q) &= p + q + \frac{1}{3}\left( p^2+q^2+p\cdot q \right)\ .
\end{align}
For effective SU(2) irreps inside SU(3) it is $n_c=0$, which yields
\begin{align}
\label{eq: p and q 2}
 p &= n_a - n_b = 2n_a-N_s\ ,\\
 q &= n_b = N_s-n_a\ .
\end{align}
We can thus write the eigenvalues of the  SU(3) quadratic Casimir operator for two-color states as
\begin{align}
 \label{eq: Quadratic casimir SU3 2}
 C_2^{\text{SU}(3)}(n_a) &= n_a + n_a^2-n_a\cdot N_s + \frac{N_s^2}{3}\ ,
\end{align}
which only depends on $N_s$ and $n_a$. We then compute the difference of the eigenvalues for increasing $n_a$
\begin{align}
 \label{eq: Casimir difference SU3}
 C_2^{\text{SU}(3)}(n_a+1)-C_2^{\text{SU}(3)}(n_a) &= 2\cdot n_a - N_s + 2 \nonumber \\
 &= p + 2\ .
\end{align}
From Eq.~\eqref{eq: Quadratic casimir SU3 2} it follows that the eigenvalues grow with $n_a$.
The smallest eigenvalue of the effective SU(2) irreps inside SU(3) is therefore given by inserting the lowest allowed value $n_a^*$, while $n_c$ remains $0$. Since the number of columns in Young tableaus must not increase going from top to bottom, this value is determined by
\begin{align}
n_a^* &= \begin{cases}
	  \frac{N_s}{2} & \text{ for }N_s\text{ even} \\
	  \frac{N_s+1}{2} & \text{ for }N_s\text{ odd}
	  \end{cases}\ .
\end{align}
The lowest eigenvalue of effective SU(2) irreps inside SU(3) is thus given by
\begin{align}
 C_2^{\text{SU}(3)}(n_a^*) &= \begin{cases}
                 \frac{N_s^2}{12}+\frac{N_s}{2} & \text{ for }N_s\text{ even} \\
	  \frac{N_s^2}{12}+\frac{N_s}{2}+\frac{3}{4} & \text{ for }N_s\text{ odd}
                \end{cases}\ ,
\end{align}
and therefore depends only on the number of sites $N_s$.
The eigenvalues of the quadratic Casimir operator of the SU(2)-like irreps of SU(3) start at $C_2^{\text{SU}(3)}(n_a^*)$ and increase by $p+2$ where $p$ starts from $0$ (or $1$ for odd $N_s$) and increases by $2$ for constant $N_s$.
Starting from SU(2), the quadratic Casimir operator only depends on $p$ since the number of rows of the corresponding Young tableaus is $2$. It can be written as
\begin{align}
 \label{eq: Quadratic casimir SU2}
 C_2^{\text{SU}(2)}(p) &= \frac{p^2}{4}+\frac{p}{2}\ ,
\end{align}
which follows by setting $S=p/2$ into the usual eigenvalue expression $C_2^{\text{SU}(2)}(S)=S(S+1)$.
For even $N_s$ the lowest eigenvalue is $0$, while for odd $N_s$ it is $3/4$.
If $p$ is increased, the eigenvalues of the SU(2) quadratic Casimir operator changes by
\begin{align}
\label{eq: Casimir difference SU2}
C_2^{\text{SU}(2)}(p+2)-C_2^{\text{SU}(2)}(p) &= p + 2\ .
\end{align}
Thus the slope is the same as for the SU(3) quadratic Casimir operator.
So, every eigenvalue of the SU(2) quadratic Casimir operator also appears in SU(3), shifted by $C_2^{\text{SU}(3)}(n_a^*)$.
E.g.~for $21$ sites, $n_a^*=11$ and $C_2^{\text{SU}(3)}(11)=48$, which perfectly agrees with Fig.~\ref{Fig:S_TOS_SU2}. Mathematically, one may summarize the situation as follows. 
Let $\mathcal{G}\in\{SU(2),SU(3)\}$ be a Lie-Group and $\sigma\left(\mathcal{G}\right)$ denote the spectrum of the Quadratic Casimir operator of the group $\mathcal{G}$, then:
\\ \newline$\forall s\in\sigma\left(SU(2)\right)\exists t\in\sigma\left(SU(3)\right)$ such that
\begin{align}
\label{eq: theorem on casimir operators}
 t-s &= \begin{cases}
	  \frac{N_s^2}{12}+\frac{N_s}{2} & \text{ for }N_s\text{ even} \\
	  \frac{N_s^2}{12}+\frac{N_s}{2}+\frac{3}{4} & \text{ for }N_s\text{ odd}
	\end{cases}\ .
\end{align}

\section{Topological properties of the CSL}
\label{sec:csltopologocalproperties}
In Sec.~\ref{sec:JKmodel}, we argue that the six-fold degenerate ground state corresponds to an Abelian CSL.
Here, we focus on the three variational states with $\pi/3$-flux and compare the numerical results from VMC with the predictions for the $\pi/3$-flux CSL.

We use the variational state overlap method~\cite{Hung2014,Moradi2015,Mei2015} to confirm the topological properties of the anyonic excitations in the CSL phase.
Similar VMC constructions were shown to produce topological fractional quantum Hall states in SU(2) systems~\cite{Lu2014, Liu2014, Hu2015b, Wietek2015}.
The information on the topological spins and mutual statistics of the anyonic quasi-particles can be determined by calculating the matrix elements of the generators of the modular transformations in the ground-state manifold on the torus~\cite{Zhang2012}.
Modular transformations are the mappings of the torus onto itself, mapping each vertex of the lattice to another one. 
All such transformations can be given as a composition of powers of two generators, $S$ and $T$ (Dehn twist), and they form the $SL(2,\mathbb{Z})$ group. 
A general modular transformation can be represented by an integer valued $2 \times 2$ matrix with unit determinant, which describes how the vectors $\vec \omega_1$ and $\vec \omega_2$ defining the torus periodicity change under the transformation.
The two generators are given by
\begin{equation}
\begin{split}
S \left( \begin{array}{c}  \vec \omega_1 \\ \vec  \omega_2\end{array}\right) &= \left( \begin{array}{cc}  0&1 \\ -1&0 \end{array}\right) \left( \begin{array}{c}  \vec \omega_1 \\  \vec\omega_2\end{array}\right)= \left( \begin{array}{c}   \vec \omega_2 \\  -\vec\omega_1\end{array}\right)\ , \\
T \left( \begin{array}{c}  \vec \omega_1 \\ \vec \omega_2\end{array}\right) &= \left( \begin{array}{cc}  1&1 \\ 0 &1 \end{array}\right) \left( \begin{array}{c}  \vec \omega_1 \\  \vec\omega_2\end{array}\right) = \left( \begin{array}{c}  \vec \omega_1+\vec \omega_2 \\  \vec \omega_2\end{array}\right)\ .
\end{split}
\end{equation}
On the square lattice the $S$ transformation is equivalent to a $\pi/2$ rotation around a site, but this is not true in general. In case of the triangular lattice $S$ is not a symmetry transformation either, but  $T^{-1}\cdot S$ is equivalent to a $\pi/3$ rotation, while $TS$ gives  a rotation by $2\pi/3$. 
 
Calculating the matrix elements of ${S}$ and ${T}$ over the ground-state manifold in the basis of minimum entropy states (MES)~\cite{Zhang2012} gives access to the topological spins and the exchange statistics of the anyonic quasi-particles
\begin{equation}
\begin{split}
\bra{\Xi_a}S \ket{\Xi_b} &= \e{-\alpha_S L^2 + \mathcal{O}(1/L^2)}\mathcal{S}_{ab}\ ,\\
\bra{\Xi_a}T \ket{\Xi_b} &= \e{-\alpha_T L^2+ \mathcal{O}(1/L^2)} \mathcal{T}_{ab}\\
&=\e{-\alpha_T L^2+ \mathcal{O}(1/L^2)}\e{-i \frac{2\pi}{24} c}\delta_{ab} \theta_a\ ,
\end{split}
\label{eq:modmxoverlap}
\end{equation}
where $\ket{\Xi_a}$ and $\ket{\Xi_b}$ are the MES~\cite{Mei2015}. The $S$ and $T$ transformations are not symmetries of the triangular lattice and they do not preserve the connectivity of the sites, therefore their matrix elements only give the desired topological quantities up to prefactors exponentially small in the system size~\cite{Mei2015}. In Eq.~\eqref{eq:modmxoverlap}, $\mathcal{S}_{ab}$ gives the phase the $a$th quasi-particle acquires when going around the $b$th quasi-particle.
In the MES basis the matrix of $T$ is diagonal and the entries of $\theta_a$ give the topological spin of each quasi-particle.
One of the topological angles is always 1 corresponding to the trivial topological sector.
The topological or chiral central charge $c$ is the difference of the central charges for the left and right moving edge excitations~\cite{Wen2012}.
%
However, the states obtained from the overlap calculations are not necessarily in the MES basis and 
diagonalizing the matrix of the $T$ transformation does not immediately give the MES basis if several quasi-particles have the same topological spin.
Therefore, some other criterion needs to be used to identify the MES basis.

Before presenting the results from our VMC calculations, we determine the expectations for the $\pi/3$-flux CSL.
On top of the trivial sector, the two other sectors correspond to two anyons with topological spins $-2\pi/3$, which we denote as $a$ and $\bar a$, emphasizing that they are the antiparticles of each other. The fusion rules are
\begin{equation}
\begin{array} {c|ccc}
\times & 1&a &\bar a\\
\hline
1 &1&a &\bar a\\ 
a & a &\bar a&1\\
\bar a& \bar a& 1&a
\end{array}\ .
\end{equation}
The Abelian nature is manifested in the fact that fusing two quasi-particles always results in only one quasi-particle.
Based on the fusion rules and the topological spins, we can deduce the elements of the $\mathcal{S}$-matrix based on the Verlinde formula~\cite{Verlinde1988}
\begin{equation}
\label{eq:verlinde}
\begin{split}
\mathcal{S}_{a b} = \sum_c  \frac{1}{D} N_{ab}^c \frac{\theta_c}{\theta_a \theta_b} d_c\ ,
\end{split}
\end{equation}
where $N_{ab}^c$ are the coefficients in the fusion rules, $d_c$ are the quantum dimensions of each quasi-particle (1 for all quasi-particles in an Abelian theory), and $D =\sqrt{\sum d_c^2}$ is the total quantum dimension. 
The topological spins in our case are $\theta_1= 1$ and $\theta_a = \theta_{\bar a} = \e{-\ci 2\pi/3}$.
The topological central charge $c$ satisfies $\e{\ci c2\pi/8}=\sum_\alpha \theta_\alpha d_\alpha^2/ D^2$ \cite{Wen1992, Wang2010}, which is consistent with $c=-2$.
The topological spin $-2\pi/3$ can be understood at the mean-field level.
In order to exchange two fermions, they need to be moved around a triangular plaquette and thus the overall phase is a combination of the $\pi$ phase of exchanging two fermions plus the extra $\pi/3$ phase picked up from the gauge flux.  Also, for the $\pi/3$-flux case the fermions go around counter-clockwise in the bulk, thus the skipping modes on the edge travel clockwise, which agrees with \mbox{$c=-2$}.
Based on this and Eq.~\eqref{eq:verlinde}, the modular matrices $\mathcal{T}$ and $\mathcal{S}$ have the form
\begin{equation}
\begin{split}
\mathcal{T}&=\e{\ci \pi/6} \left(\begin{array}{ccc} 1&0&0\\0 &\e{-\ci 2\pi/3}&0\\0&0& \e{-\ci 2\pi/3}\end{array}\right)\ , \\
\mathcal{S}&= \frac{1}{\sqrt{3}}\left(\begin{array}{ccc} 1&1&1\\1 &\e{\ci 2\pi/3}&\e{-\ci 2\pi/3}\\1&\e{-\ci 2\pi/3}& \e{\ci 2\pi/3}\end{array}\right)\ ,
\end{split}
\label{eq:modmxprediction}
\end{equation}
and since it is an Abelian spin liquid all elements of $\mathcal{S}$ have the same magnitude.

\begin{figure}[t]
\begin{center}
\includegraphics[width=0.45\textwidth]{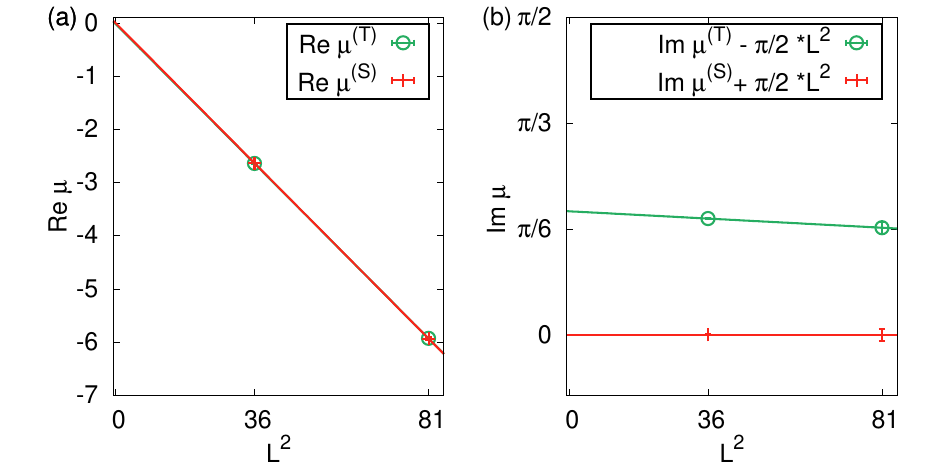}
\caption{Scaling of the prefactors $\mu^{(S)}$ and $\mu^{(T)}$ as a function of the squared system size $L^2$. We show data for $L=\{6, 9\}$. The extrapolated value of $\Im \mu^{(T)}$ gives an estimate on the chiral central charge of $c\approx -2.34$.
The value $\Im \mu^{T}(L=0)= \pi/6$ would correspond to central charge $c=-2$.}
\label{fig:prefactor}
\end{center}
\end{figure}

Next, we present our VMC results for the matrix elements of the $S$ and $T$ modular transformations and show that they agree with the predictions for the Abelian CSL phase.
The modular matrices are based on $10^{10}$ and $10^{9}$ independent measurements in 100 batches for the $6\times 6$ and $9\times 9$ clusters, respectively.
The MES basis is identified with the case where, on top of the $T$-matrix being diagonal, all elements of the $S$-matrix have the same magnitude.
In order to fix the overall phase of the minimally entangled states, one can use the fact that the row and column of the $S$ matrix corresponding to the trivial sector should be $1/\sqrt{3}$.
This yields
\begin{widetext}
\begin{equation}
\begin{split}
T_{(36)}&= \e{ \mu^{(T)}_{(36)}}\left(
\begin{array}{ccc}
 1& 0& 0\\
0 & 1.005 \e{-\ci 0.699 \pi}& 0\\
0 & 0& 1.017\e{- \ci 0.683 \pi} 
\end{array}
\right),
S_{(36)}= \frac{\e{ \mu^{(S)}_{(36)}}}{\sqrt{3}}
\left(\begin{array}{ccc}
 0.965& 1.006 & 1.006 \\
 1.006 & 1.003 \e{ \ci 0.663\pi}& 1.003 \e{-\ci 0.663 \pi} \\
 1.006 & 1.003 \e{ -\ci 0.663\pi}& 1.003 \e{\ci 0.663\pi}
\end{array}
\right),\\
T_{(81)} &= \e{ \mu^{(T)}_{(81)}}\left(
\begin{array}{ccc}
 1& 0 &0\\
0& 0.991\e{-\ci 0.668 \pi}  &0\\
0  & 0 & 1.005 \e{-\ci 0.667\pi}
\end{array}
\right),
S_{(81)}=\frac{\e{ \mu^{(S)}_{(81)}}}{\sqrt{3}}
\left(
\begin{array}{ccc}
 1.018 & 1.009 & 0.990 \\
 1.015 & 0.997 \e{\ci 0.676 \pi} & 1.002 \e{-\ci 0.670 \pi} \\
 0.984 & 0.989 \e{-\ci 0.664 \pi} & 0.995 \e{\ci 0.658 \pi}
\end{array}
\right).
\end{split}
\label{eq:STmx}
\end{equation}
\end{widetext}
for the $6 \times 6$- and $9 \times 9$-site system, respectively.

In the left panel of Fig.~\ref{fig:prefactor}, we show how the real part of the prefactors $\Re \mu^{(S)}$ and $\Re \mu^{(T)}$ scale with the system size.
Our results fulfill an $L^2$ scaling, and we find an estimate for the chiral central charge $c \approx - 2.34$, which  is consistent with the expected $c=-2$.
In the right panel of Fig.~\ref{fig:prefactor}, we present the imaginary part of the prefactors $\Im \mu^{(S)}$ and $\Im \mu^{(T)}$ corrected by the phase factors $\pm \pi/2 \cdot L^2$, which does not change the extrapolated value.
We note that the values of $\Im \mu^{(S)}$ and $\Im \mu^{(T)}$ can be extracted only mod $2\pi$.
So the values for the different system sizes could be shifted by an integer multiple of $2\pi$ independently of each other.
This can change the extrapolated value at $L=0$ by an integer multiple of $2\pi/5$, which would result in a shift of an integer multiple of $4.8$ in the extrapolated chiral central charge $c$.
So, this is clearly not the source of the discrepancy between the extrapolated and theoretical values. 

\begin{table}[b!]
\begin{center}
\begin{tabular}{|l|l|l|}
\hline
 &  $6\times 6$ & $9\times9 $\\
\hline
$\Delta|T_{ab}| $ & $\pm 0.0005 $ & $\pm 0.004 $ \\
\hline  
$\Delta |S_{ab} |$ & $\pm 0.005 $ & $\pm 0.02$ \\
\hline
$\Delta \arg(T_{aa})$ & $\pm 0.0001 \pi $ & $\pm0.01\pi$ \\
\hline
$\Delta\arg(S_{ab})$ & $\pm 0.005 \pi$ &$ \pm0.05\pi  $\\
\hline
\end{tabular}
\end{center}
\caption{Estimates of the numerical/stochastic error of the matrix elements of the numerically obtained modular matrices from VMC.}
\label{tab_errors}
\end{table}

We used the results from independent batches of calculations (compare Sec.~\ref{sec:methodsVMC}) to estimate the statistical errors of the simulation.
These are collected in Tab.~\ref{tab_errors}.
However, the main source of error is systematic, and stems from the finite system sizes as well as from the fact that the projected states are not perfect topological CSL states.
%
%
Carrying out calculations on larger systems could help to further verify our results, but based on the calculations on the 36- and 81-site clusters we find that for the $12 \times 12$ system the exponential prefactor would be $\mathcal{O}(10^{-5})$.
Since the error of Monte Carlo measurements scales as the inverse square root of the number of samples, it would require a minimum of $10^{12}$ samples to see anything beyond the numerical error.
This goes beyond the scope of our current VMC method.

\section{Symmetries of the CSL}
\label{sec:CSL:symmetries}
\begin{table}[b!]
\centering
\begin{tabular}{|c|c|c|c|c|c|c|}
  \hline
   & $\ket{\Psi_1}$ & $\ket{\Psi_2}$ & $\ket{\Psi_3}$ & $\ket{\Psi_4}$ & $\ket{\Psi_5}$ & $\ket{\Psi_6}$ \\ \hline \hline
  $t_1$ & 1 & $\e{-i2\pi/3}$ & $\e{i2\pi/3}$ & 1 & $\e{-i2\pi/3}$ & $\e{i2\pi/3}$ \\ \hline
  $t_2$ & 1 & $\e{i2\pi/3}$ & $\e{-i2\pi/3}$ & 1 & $\e{i2\pi/3}$ & $\e{-i2\pi/3}$ \\ \hline
  \hline
   $\phantom{a}$ & $\ket{\Psi_1^{\phantom{*}}}$ & $\ket{\Psi_2^*}$ & $\ket{\Psi_3^*}$ & $\ket{\Psi_4^{\phantom{*}}}$ & $\ket{\Psi_5^*}$ & $\ket{\Psi_6^*}$ \\ \hline \hline
  $r_{\pi/3}$ & 1 & $\e{-i2\pi/3}$ & $\e{i\pi/3}$ & 1 & $\e{i2\pi/3}$ & $\e{-i\pi/3}$ \\ \hline
\end{tabular}
\caption{Eigenvalues of symmetry operators for the six chiral states $\ket{\Psi_i}$ or $\ket{\Psi_i^*}$ ($\ket{\Psi_1}$ is the ground state, $\ket{\Psi_2}$ and $\ket{\Psi_3}$ are the first-excited states, etc.) on the 12-site cluster. The states $\ket{\Psi_2}$ and $\ket{\Psi_3}$ as well as $\ket{\Psi_5}$ and $\ket{\Psi_6}$ are degenerate.
The eigenvalues $t_1$($t_2$) correspond to a translation along the lattice vectors $\vec{r}_1$($\vec{r}_2$). The eigenstates denoted without (with) a star are diagonalized in the joint eigenbasis of the Hamiltonian and the translation (rotation) operator. The results from ED and VMC are identical.}
\label{Tab:symmetries12}
\end{table}

\begin{table}[b!]
\centering
\begin{tabular}{|c|c|c|c|c|c|c|}
  \hline
   & $\ket{\Psi_1}$ & $\ket{\Psi_2}$ & $\ket{\Psi_3}$ & $\ket{\Psi_4}$ & $\ket{\Psi_5}$ & $\ket{\Psi_6}$ \\ \hline \hline
  $t_1$ & 1 & $\e{-i2\pi/3}$ & $\e{i2\pi/3}$ & 1 & $\e{-i2\pi/3}$ & $\e{i2\pi/3}$ \\ \hline
  $t_2$ & 1 & $\e{i2\pi/3}$ & $\e{-i2\pi/3}$ & 1 & $\e{i2\pi/3}$ & $\e{-i2\pi/3}$ \\ \hline
  \hline
   $\phantom{a}$ & $\ket{\Psi_1^{\phantom{*}}}$ & $\ket{\Psi_2^*}$ & $\ket{\Psi_3^*}$ & $\ket{\Psi_4^{\phantom{*}}}$ & $\ket{\Psi_5^*}$ & $\ket{\Psi_6^*}$ \\ \hline \hline
  $r_{\pi/3}$ & -1 & $\e{-i2\pi/3}$ & $\e{i\pi/3}$ & -1 & $\e{i2\pi/3}$ & $\e{-i\pi/3}$ \\ \hline
\end{tabular}
\caption{Eigenvalues of symmetry operators for the six chiral states on the 21-site cluster.
The notation is as in Tab.~\ref{Tab:symmetries12}. The degeneracies on 12 and 21 sites are identical, as are the results from ED and VMC.}
\label{Tab:symmetries21}
\end{table}

\begin{table}[t!]
\centering
\begin{tabular}{|c|c|c|c|c|c|c|}
  \hline
   & $\ket{\Psi_1}$ & $\ket{\Psi_2}$ & $\ket{\Psi_3}$ & $\ket{\Psi_4}$ & $\ket{\Psi_5}$ & $\ket{\Psi_6}$ \\ \hline \hline
  $t_1$,  $t_2$  & 1 & 1&1&1&1&1 \\ \hline
  $r_{\pi/3}$ & 1 & 1 & $\e{-i2\pi/3}$ &$\e{i2\pi/3}$ & -1 & -1\\ \hline
  irrep of $D_6$ & $A_1$ & $A_2$ &\multicolumn{2}{c|}{$ E_2$} & $B_2$& $B_1$ \\ \hline
\end{tabular}
\caption{Eigenvalues of symmetry operators for the six chiral states on the 36-site cluster from VMC only. The notation is similar to Tab.~\ref{Tab:symmetries12}. On 36 sites all the 6 states are at the $\Gamma$ point in the Brillouin zone. The states $\ket{\Psi_3}$ and $\ket{\Psi_4}$ are degenerate, the others are not.}
\label{Tab:symmetries36}
\end{table}

Let the lattice constant be $a$ and the lattice vectors \mbox{$\vec{r}_1=(a,0)$} and $\vec{r}_2=a(\cos(\pi/3),\sin(\pi/3))$, and let us denote by $T_1$ and $T_2$ the corresponding translation operators. Furthermore, let us define the rotation operator $R_{\pi/3}$ that rotates counterclockwise by an angle $\pi/3$. These translations and the rotation do not have a joint eigenbasis. To distinguish them, we will mark the eigenstates of $H$ and $R_{\pi/3}$ with a star.

The eigenvalues for the six chiral states on the 12- and 21-site clusters are given in Tab.~\ref{Tab:symmetries12} and \ref{Tab:symmetries21}, respectively. The results from the $\pm \pi/3$ CSL variational states from VMC and from the low-lying energy eigenstates at $K/J=0.35$ ($\alpha \approx 0.11 \pi$) from ED show a perfect match. 

The spectrum of the 12-site cluster is discussed in Sec.~\ref{sec:hubbardcomparison}. On this particular system the ground-state manifold of six states is intertwined with another state. However, if the states are analyzed with respect to their symmetry values, the apparent six low-lying CSL states can be identified.

The symmetry properties of the chiral states for clusters larger than 21 sites were only studied by VMC. For systems where the vectors defining the torus lie in the $\vec{r}_1$ and $\vec{r}_2$ directions all six chiral states have a wave vector zero. The smallest example is the 36-site cluster for which we provide the symmetry properties in Tab.~\ref{Tab:symmetries36}. For this specific finite size, only one pair of states is degenerate, the other states are not.

\section{Effective model on 12-site cluster with PBCs}
\label{sec:apptwelvesites}
\begin{figure}[b]
\centering
\includegraphics[width=0.35\columnwidth, angle = 0, trim={5cm 9cm 4cm 2.5cm},clip]{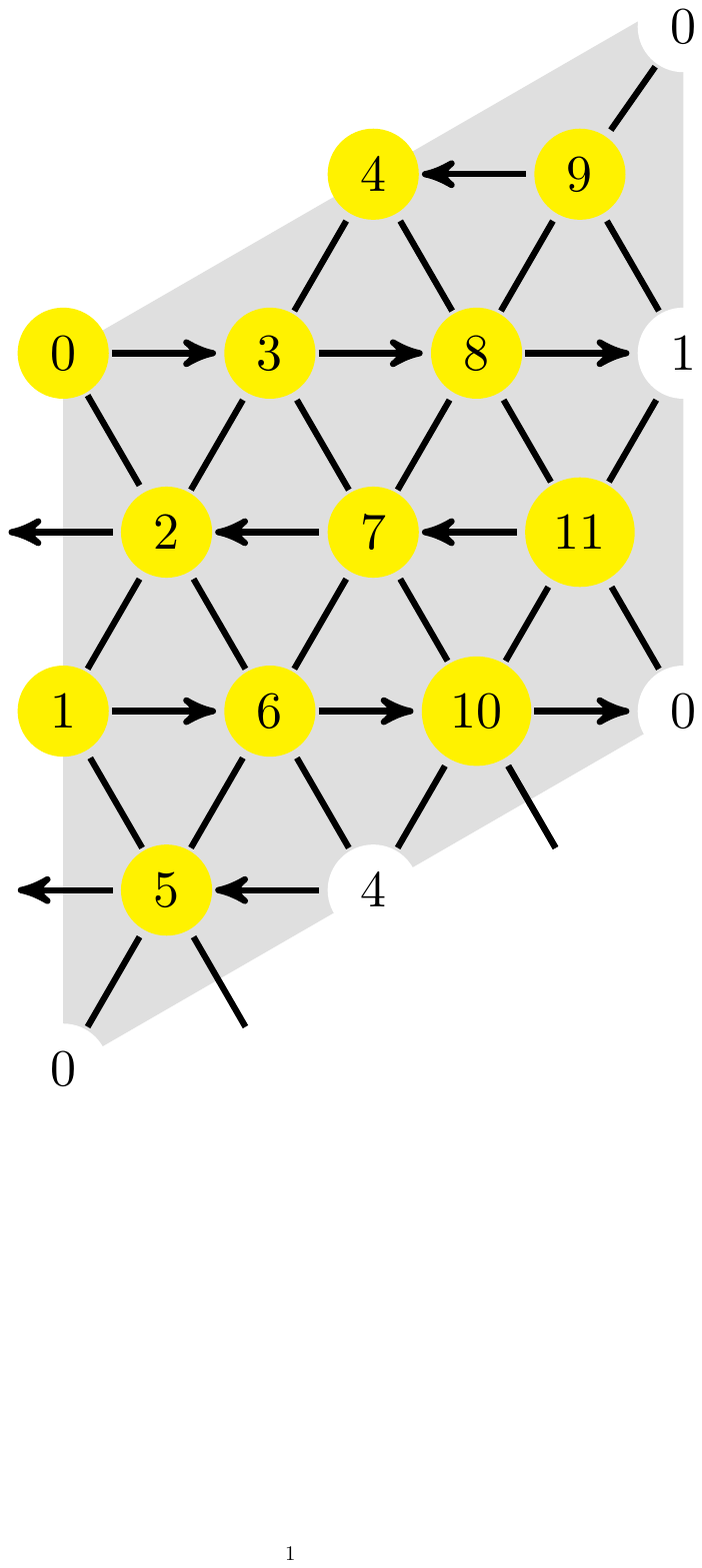}
\caption{Triangular lattice cluster with 12 sites (yellow) and PBCs (white). The chosen gauge is illustrated by the arrows on the bonds: A hopping in the direction of the arrow (resp. in the opposite direction) has a positive phase factor $\e{\ci\Phi}$ (resp. the complex conjugate $\e{-\ci\Phi}$.}
\label{Fig_12siteclusterwithgauge}
\end{figure}

On a cluster consisting of a finite number of sites the effective model is different from the model in the thermodynamic limit. In the case of PBCs this is due to the fact that a finite number of fermionic hoppings in one direction leads back to the starting site. For the 12-site torus this becomes relevant in order 4 in $t/U$, where the four-site plaquette can be embedded surrounding the cluster via the PBCs.
This leads to an additional effective interaction around the 12-site torus, but also to modifications of other coupling constants compared to the infinite lattice.
%
The easiest approach to derive the effective Hamiltonian for the most interesting case $\Phi = \pi$ is to consider the model without phases and then perform the transformation $t \rightarrow -t$, which is identical to $\Phi=0 \rightarrow \Phi=\pi$.
For general fluxes $\Phi$ the calculation is done by choosing the gauge for the 12-site cluster depicted in Fig.~\ref{Fig_12siteclusterwithgauge}.
Every second line of horizontal bonds in the cluster gets assigned with a phase $\Phi$, whereas the parallel intermediate bonds contribute with a phase $-\Phi$. All non-parallel bonds yield a vanishing phase.
With this gauge some of the exchange constants in the effective model on 12 sites, which are symmetric on the infinite lattice, take different values along different directions.
For instance, the nearest-neighbor exchange on the bonds with a phase $J_{\text{horiz}}$ and the nearest-neighbor exchange on the bonds without a phase $J_{\text{diag}}$ differ.
From the perspective of a LCE this relates to the contributions from the four-site plaquette looping around the torus with distinct phases for different directions.
In order 4 in $t/U$ we find
\begin{equation}
\label{eq_effmodel12sites_J}
\begin{aligned}
&J_{\text{hor}} = 2 \frac{t^2}{U^2} + 12 \cos\Phi\frac{t^3}{U^3} + \left( 60+40\cos 2\Phi\right) \frac{t^4}{U^4}\ ,\\
&J_{\text{dia}} =  2 \frac{t^2}{U^2} + 12 \cos\Phi\frac{t^3}{U^3} + \left( 36+64\cos 2\Phi\right) \frac{t^4}{U^4}\ .
\end{aligned}
\end{equation}
For the newly arising four-site ring exchange around the PBCs different phases occur depending on the location.
In total there are three directions with $18$ loops each. All ring exchanges along the vertical direction have a zero flux. For the two other directions $2/3$ of the exchanges contribute with a phase factor, whereas $1/3$ of them do not.
Similarly, for the constant part \mbox{$E_0=12\epsilon_0$} several versions of the 4-site plaquettes either with or without fluxes have to be taken into account.
The full effective Hamiltonian can be written
\begin{widetext}
\begin{equation}
\label{Eq_H_eff12sites_triangular}
\begin{split}
\mathcal{H}&_\text{triangular,\text{12sites},pbc}^{\mathcal{O}\left( 4\right)}=
12 \epsilon_0
+ J_{\text{hor}} \sum_{\includegraphics[width=0.042\linewidth, trim={3cm 27cm 16cm 1.9cm},clip]{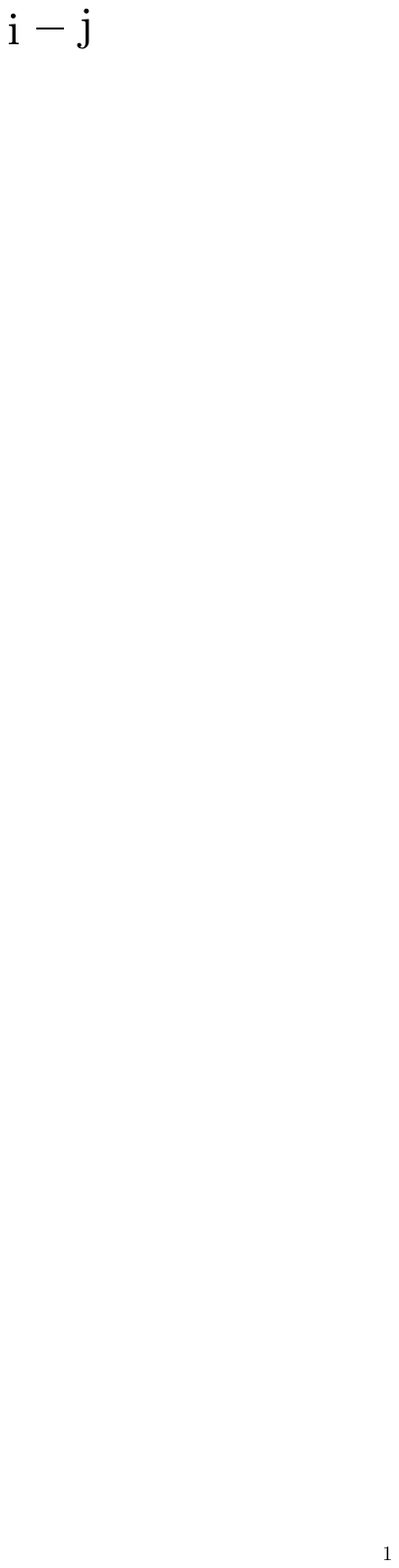}} P_{ij}
+ J_{\text{dia}} \sum_{\includegraphics[width=0.06\linewidth, trim={2.7cm 25.7cm 15cm 2cm},clip]{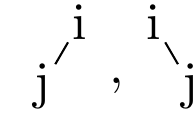}} P_{ij}
+ \sum_{\includegraphics[width=0.042\linewidth, trim={3cm 26cm 16cm 1.9cm},clip]{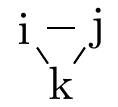}}
\left( K P_{ijk} + \text{h.c.} \right)
\\
&+ \frac13 L_{\text{s}}^{\text{2sp}}\sum_{\includegraphics[width=0.055\linewidth, trim={3cm 27cm 15cm 1.9cm},clip]{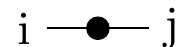}} P_{ij}
+ \frac12 L_{\text{d}}^{\text{2sp,diag}}\sum_{\includegraphics[width=0.05\linewidth, trim={3cm 26cm 15.5cm 1.9cm},clip]{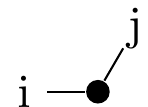}} P_{ij}
+ \frac12 L_{\text{d}}^{\text{2sp,vert}} \sum_{\includegraphics[width=0.06\linewidth, trim={2.5cm 24.7cm 15cm 2cm},clip]{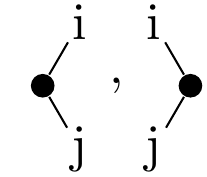}} P_{ij}
\\
& + \sum_{\includegraphics[width=0.1\linewidth, trim={2cm 24.7cm 13cm 2cm},clip]{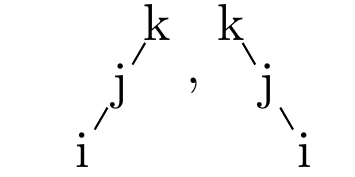}} \left( L_{\text{s,diag}}^{\text{3sp}} P_{ijk} + \text{h.c.} \right)
+ \sum_{\includegraphics[width=0.065\linewidth, trim={3cm 26cm 14.7cm 1.9cm},clip]{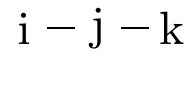}}
\left(  L_{\text{s,horiz}}^{\text{3sp}}  P_{kji} + \text{h.c.} \right)
\\
& + \sum_{\includegraphics[width=0.1\linewidth, trim={3cm 25.6cm 12cm 2cm},clip]{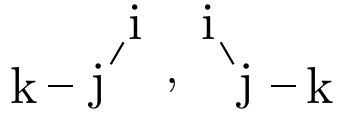}} \left( L_{\text{d}}^{\text{3sp,diag}} P_{ijk} + \text{h.c.} \right)
+ \sum_{\includegraphics[width=0.1\linewidth, trim={3cm 25.6cm 12cm 2cm},clip]{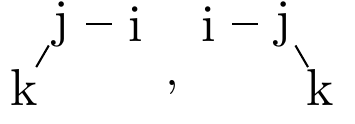}} \left( L_{\text{d}}^{\text{3sp,diag}} P_{kji} + \text{h.c.} \right)
\\
&+\sum_{\includegraphics[width=0.055\linewidth, trim={3cm 24.9cm 15cm 2cm},clip]{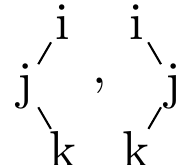}} \left( L_{\text{d}}^{\text{3sp,vert}} P_{kji} + \text{h.c.} \right)
+ L_{\text{r}}^{\text{4sp},\Phi=0}
\sum_{\includegraphics[width=0.04\linewidth, trim={3cm 24.9cm 16cm 2cm},clip]{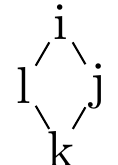}}
\left( P_{lkji} + \text{h.c.} \right)
\\
&+ \sum_{\includegraphics[width=0.1\linewidth, trim={3cm 25.5cm 13cm 2cm},clip]{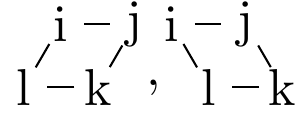}}
\left( L_{\text{r}}^{\text{4sp},\Phi\neq 0} P_{ijkl} + \text{h.c.} \right)
+ \sum_{\text{torus loops}} \left( L_{\text{rpbc}}^{\text{4sp}}(\Phi  P_{ijkl} + \text{h.c.} \right),
\end{split}
\end{equation}
\end{widetext}
where the depicted graphs under every sum indicate in which orientations the exchanges lie on the cluster. For the three site ring exchange on a triangle $K$ all triangles have to be taken. The four-site ring exchange around the torus is only stated. The explicit shapes differ and lead to either a phase factor or not; In total every site is part of 18 four-site ring exchanges around the PBCs, wherein 10 contribute without a phase factor and 8 with a phase factor. In total these exchanges double due to the hermitian conjugated exchanges.
For the real exchange constants ($J, L_{\text{s}}^{\text{2sp}}, L_{\text{d}}^{\text{2sp}}, L_{\text{d, vert}}^{\text{3sp}}, L_{\text{r, pbc},\Phi=0}^{\text{4sp}}$) the turning direction is arbitrary, since the Cosine is symmetric and the coupling constants are identical under the transformation $\Phi \rightarrow - \Phi$.
For the exchanges contributing with a non-trivial phase factor the turning direction has to be chosen accordingly. For 2-site interactions such a turning direction cannot be defined, which is why such couplings contribute with purely symmetric phase factors.

The effective nearest-neighbor exchange constants are given in Eq.~\eqref{eq_effmodel12sites_J}. All other couplings and the constant contributions are
\begin{equation}
\begin{aligned}
&K = - 6 \e{\ci\Phi} \frac{t^3}{U^3} + \left(- 14 - 20 \e{2\ci \Phi} \right) \frac{t^4}{U^4}\ ,\\
&L_{\text{s}}^{\text{2sp}} = \left( 18 + 16 \cos 2\Phi  \right) \frac{t^4}{U^4}\ ,\\
&L_{\text{d,diag}}^{\text{2sp}} = \left( \frac{88}{3} + 32 \cos 2\Phi  \right) \frac{t^4}{U^4}\ ,\\
&L_{\text{d,vert}}^{\text{2sp}} = \frac{184}{3} \frac{t^4}{U^4}\ ,\\
&L_{\text{s,diag}}^{\text{3sp}} = \left( -\frac{34}{3} -10 \e{2 \ci \Phi} \right) \frac{t^4}{U^4}\ ,\\
&L_{\text{s,horiz}}^{\text{3sp}} = \left( -\frac{4}{3} -20 \e{2 \ci \Phi} \right) \frac{t^4}{U^4}\ ,\\
&L_{\text{d,diag}}^{\text{3sp}} = \left(-\frac{34}{3} - 20 \e{2 \ci \Phi} \right) \frac{t^4}{U^4}\ ,\\
&L_{\text{d,vert}}^{\text{3sp}} = -\frac{94}{3} \frac{t^4}{U^4}\ ,\\
&L_{\text{r}}^{\text{4sp}}= 20 \e{2 \ci \Phi} \frac{t^4}{U^4}\ ,\\
&L_{\text{r,pbc},\Phi\neq 0}^{\text{4sp}} = 20 \e{2 \ci \Phi} \frac{t^4}{U^4}\ ,\\
&L_{\text{r,pbc},\Phi = 0}^{\text{4sp}} = 20 \frac{t^4}{U^4}\ ,\\
&\epsilon_0 = -6 \frac{t^2}{U^2} - 12 \cos\Phi \frac{t^3}{U^3}  - \left(54+32 \cos 2\Phi \right) \frac{t^4}{U^4}\ .
\end{aligned}
\end{equation}

Note that one could also derive an effective model for the 12-site cluster in which neither phases around the torus nor distinct exchange constants for topologically equivalent interactions (like for $J_{\text{hor}}$ and $J_{\text{dia}}$) occur. This can be achieved by fulfilling the $\Phi$-flux condition on every triangle without assigning specific phases to specific bonds. However, this does not matter since the eigenenergies do not depend on the gauge.

\section{CSL for $\Phi \neq \pi$}
\label{sec:CSLnonpiflux}
For the $J$-$K$ model at $\Phi=\pi/2$, hence for purely imaginary ring exchange, a $\pi/3$-flux CSL was discovered for SU(3)-symmetric spins in Ref.~\onlinecite{Nataf2016}.
As for the spontaneous time-reversal symmetry breaking CSL found for the same model but at $\Phi=\pi$, discussed in the main part of the manuscript, the relevant question is whether this CSL is a feature of the Hubbard model, and therefore reachable in experiments with artificial gauge fields.

\begin{figure}[t]
\centering
\includegraphics[width=1\columnwidth]{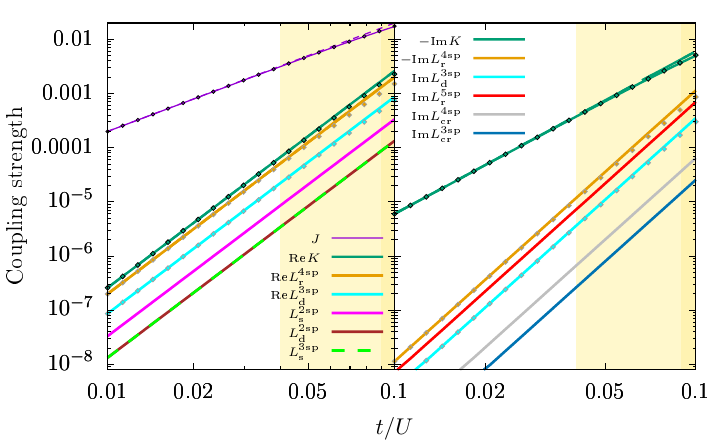}
\caption{Effective couplings in units of $U$ of the Hubbard model as a function of $t/U$ for $\Phi=\pi/2$ in a double logarithmic plot. Shown are the non-vanishing real (left) and imaginary (right) contributions in bare orders. The diamonds encircled in black (gray) give the Pad\'e extrapolants with the exponents [3,2] ([2,1]). Note that all uneven (even) terms in the real (imaginary) part vanish. The background colors are defined as in Fig.~\ref{Fig:S_coupling_convergence_pi} in the main text.}
	\label{Fig:S_couplings_piover2}
\end{figure}

\begin{figure}[t]
\centering
\includegraphics[width=1\columnwidth]{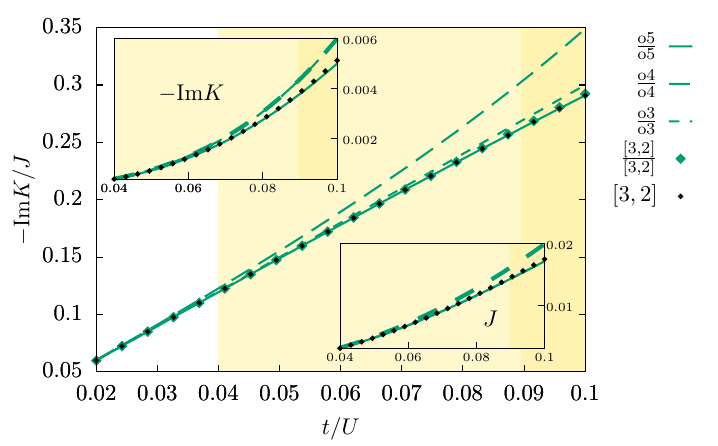}
\caption{Imaginary part of the ratio of effective coupling constants $-K/J$ depending on $t/U$ for $\Phi=\pi/2$ using bare series up to order 5. The ratios of Pad\'e extrapolants with the exponents [3,2] as well as the direct Pad\'e extrapolation of the ratio $-\text{Im}K/J$ are indicated. The insets show similar plots for the imaginary part of the negative three-site ring exchange $-\Im K$ and the nearest-neighbor exchange $J$. Extrapolations of $J$ with different pairs of exponents [2,2] and [2,3] are identical at $\Phi=\pi/2$. The background colors are defined as in Fig.~\ref{Fig:S_coupling_convergence_pi} in the main part of the manuscript.}
\label{Fig:S_K_conv_piover2}
\end{figure}

We first study the fifth-order effective model in Eq.~\eqref{eq:H_eff_triangular} for $\Phi = \pi/2$ and find that the same set of Pad\'e extrapolations as described for $\Phi=\pi$ in Sec.~\ref{sec_hubbard_effmodels} works best.
The convergence behavior of all coupling constants is shown in Fig.~\ref{Fig:S_couplings_piover2}.
%
%
%
%
A number of interesting and subtle features of the effective model become clear.
The three-site ring exchange $K$ is purely imaginary only in order 3. The fourth-order term partly arises from fluctuations around two triangles leading to a flux of $2\Phi$, therefore a real part is present in higher orders. Similarly, the imaginary part of the fourth-order contribution to the four-site ring exchange vanishes and it effectively becomes an order 5 term. As a consequence the model is dominated by a real nearest-neigbor and an imaginary three-site ring exchange. For exactly this subset of interactions the CSL phase occurs for $\Im(K/J) \gtrsim 0.3$ ~\cite{Nataf2016}.
In Fig.~\ref{Fig:S_K_conv_piover2} the ratios of Pad\'e extrapolants for the imaginary part of $-K$ and the real amplitude $J$ (green larger diamonds) and the direct Pad\'e extrapolation of the ratio $-\text{Im}K/J$ (black smaller diamonds) are shown. We see that in the regime of the CSL not only the extrapolations, but also the bare fifth-order (solid line) series are well converged.

%
%

Second, the numerical study of the full fifth-order effective model yields the signature of the CSL phase by three low-lying chiral states.
Using ED on the 21-site cluster we find $(t/U)^{\mathcal{O}(3)\text{,ED}}_\text{c}\approx (t/U)^{\mathcal{O}(5)\text{,ED}}_\text{c} \approx 0.09$ for the third- and fifth-order effective model. In VMC the values are smaller $(t/U)^{\mathcal{O}(3)\text{,VMC}}_\text{c}\approx (t/U)^{\mathcal{O}(5)\text{,VMC}}_\text{c}\approx 0.04$. This is plausible since the VMC captures CSL phases more naturally than long-range ordered phases. The apparent areas in Fig.~\ref{Fig:S_couplings_piover2} and Fig.~\ref{Fig:S_K_conv_piover2} are shaded in yellow.

This CSL is stable in the $J$-$K$ model for fluxes \mbox{$\pi/2 \leq \Phi \leq \pi$}.
The fifth-order effective model in this range partly suffers from spurious poles in the Pad\'e extrapolants of several coupling constants.
More precisely, for $L_{\text{d}}^{\text{2sp}}$ around $\Phi\approx 0.6\pi$ and for $L_{\text{s}}^{\text{3sp}}$ around $\Phi\approx 0.55\pi$ and $\Phi\approx 0.75\pi$ divergences occur, depending on $t/U$. Here one has to use the bare series, which generally decreases the quality of convergence.

Furthermore, similar to $\Phi=\pi/2$, cancellation effects of different orders at specific values of the flux $\Phi$ lead to additional subtleties in the convergence behavior. For most parts of the full $\Phi$ phase diagram we find a good convergence by comparing the ED and VMC results in different orders. It is only in the area around $\Phi\approx 3\pi/4$ that the results between fourth- and fifth-order change qualitatively.
A possible explanation is the strong suppression of a lower-order term compared to a higher-order one for certain values of $\Phi$ and $t/U$, which can lead to a decreased quality of convergence.
Here the real part of the leading fourth-order ring exchange vanishes in order 5 due to an alternating behavior by contrast to a large monotonic five-site ring-exchange. The interplay of these two effects arises only in order 5.

In summary, the $\pi/3$-flux CSL phase at $\Phi = \pi/2$ is extended in the $J$-$K$ model for larger fluxes $\Phi$ up to the spontaneous time-reversal symmetry broken case at $\Phi = \pi$.
At $\Phi = \pi/2$ the CSL is a robust feature within the effective fifth-order model describing the SU(3) Hubbard model in the strong-coupling regime.
If the metal-insulator transition occurs at values similar to the ones at $\Phi = \pi$, the $\pi/3$-flux CSL is also a feature of the SU(3) Hubbard model at $\Phi = \pi/2$ on the triangular lattice.
Whether this phase is directly connected to the spontaneous time-reversal symmetry broken CSL present at $\Phi=\pi$ remains an open question.

\bibliography{SU3_CSL}

\end{document}